\documentclass[%
 aip,
 amsmath,amssymb,
 reprint,%
]{revtex4-1}

\usepackage{graphicx}
\usepackage{dcolumn}
\usepackage{bm}

\usepackage{color}
\usepackage[utf8]{inputenc}
\usepackage[T1]{fontenc}
\usepackage{mathptmx}
\usepackage{etoolbox}

\makeatletter
\def\@email#1#2{%
 \endgroup
 \patchcmd{\titleblock@produce}
  {\frontmatter@RRAPformat}
  {\frontmatter@RRAPformat{\produce@RRAP{*#1\href{mailto:#2}{#2}}}\frontmatter@RRAPformat}
  {}{}
}%
\makeatother

\begin{document}

\preprint{APS/123-QED}

\title{Cryogenic hybrid magnonic circuits based on spalled YIG thin films}

\author{Jing Xu}
\thanks{J. Xu and C. Horn contributed equally to this work.}
\affiliation{ 
    Pritzker School of Molecular Engineering, University of Chicago, Chicago, IL 60637, USA
}
\affiliation{ 
    Center for Nanoscale Materials, Argonne National Laboratory, Lemont, IL 60439, USA
}
\affiliation{ 
    Department of Physics, University of Central Florida, Orlando, FL 32816, USA
}

\author{Connor Horn}
\thanks{J. Xu and C. Horn contributed equally to this work.}
\affiliation{ 
    Pritzker School of Molecular Engineering, University of Chicago, Chicago, IL 60637, USA
}
\author{Yu Jiang}
\affiliation{ 
    Department of Electrical and Computer Engineering, Northeastern University, Boston, MA 02115, USA
}

\author{Amin Pishehvar}
\affiliation{ 
    Department of Electrical and Computer Engineering, Northeastern University, Boston, MA 02115, USA
}

\author{Xinhao Li}
\affiliation{ 
    Center for Nanoscale Materials, Argonne National Laboratory, Lemont, IL 60439, USA
}
\author{Daniel Rosenmann}
\affiliation{ 
    Center for Nanoscale Materials, Argonne National Laboratory, Lemont, IL 60439, USA
}
\author{Xu Han}
\affiliation{ 
    Center for Nanoscale Materials, Argonne National Laboratory, Lemont, IL 60439, USA
}
\author{Miguel Levy}
\affiliation{ 
    Department of Physics, Michigan Technological University, Houghton, MI 49931, USA
}
\author{Supratik Guha$^{\dagger}$}
\email{guha@uchicago.edu}
\affiliation{ 
    Pritzker School of Molecular Engineering, University of Chicago, Chicago, IL 60637, USA
}

\author{Xufeng Zhang$^{\ddag}$}
\email{xu.zhang@northeastern.edu}
\affiliation{ 
    Department of Electrical and Computer Engineering, Northeastern University, Boston, MA 02115, USA
}
\affiliation{ 
    Department of Physics, Northeastern University, Boston, MA 02115, USA
}

\date{\today}

\begin{abstract}
Yttrium iron garnet (YIG) magnonics has garnered significant research interest because of the unique properties of magnons (quasiparticles of collective spin excitation) for signal processing. In particular, hybrid systems based on YIG magnonics show great promise for quantum information science due to their broad frequency tunability and strong compatibility with other platforms. However, their broad applications have been severely constrained by substantial microwave loss in the gadolinium gallium garnet (GGG) substrate at cryogenic temperatures. In this study, we demonstrate that YIG thin films can be spalled from YIG/GGG samples. Our approach is validated by measuring hybrid devices comprising superconducting resonators and spalled YIG films, which exhibits anti-crossing features that indicate strong coupling between magnons and microwave photons. Such new capability of separating YIG thin films from GGG substrates via spalling, and the integrated superconductor-YIG devices represent a significant advancement for integrated magnonic devices, paving the way for advanced magnon-based coherent information processing.
\end{abstract}


\maketitle


\section{Introduction}

The field of YIG magnonics \cite{Serga2010Jun} is a rapidly evolving research area focusing on the study of collective spin excitations (magnons) in YIG (yttrium iron garnet) crystals. In recent years, magnonics has shown great potential for hybrid information processing \cite{Awschalom_IEEETransQuantEng_2021,YiLi_JAP_2020,Zhang2023Sep,Lachance_APE_2019}. As a ferrimagnetic insulator, YIG has been considered as one of the most attractive magnetic materials for developing hybrid magnonic devices, thanks to its low magnetic damping, high spin density, and excellent compatibility with various physical systems such as microwave, optics, and acoustics. These unique properties enable the extensive exploration in recent years on YIG-based hybrid magnonic devices such as electromagnonics \cite{XufengPRL2014,Tabuchi2014Aug,LihuiPRL2015,Goryachev2014Nov,Huebl2013Sep,Harder_SSC_2018,Bhoi_SSP_2020,Hu2020Jan,Rameshti_PhysRep_2022,Rao2019Jul,Xu2021May}, optomagnonics \cite{Zhang2016Sep,Bi2011Dec,Zhu2020Oct}, and magnomechanics \cite{Seo2017Mar,An2020Feb,Zhang2016Mar,Xu2021Aug} for different applications ranging from quantum information science \cite{Tabuchi2015Jul,Lachance-Quirion2020Jan,Lachance-Quirion2017Jul,Wolski2020Sep,Xu2023May} to dark matter detection \cite{Crescini_PRL_2020,Flower_PhysDarkUniv_2019}. With the recent growing interests in developing large-scale integrated magnonic circuits, thin-film YIG devices are now highly sought after, marking a shift from the bulk YIG spheres widely used in earlier research.

However, the development of thin film YIG devices at cryogenic temperatures has faced significant limitations, primarily due to the undesirable properties of the substrate used for YIG thin film growth. The optimal method for producing high-quality crystalline YIG films is epitaxial growth on gadolinium gallium garnet (GGG) substrates, which have a lattice constant closely matched to that of YIG, resulting in room-temperature magnon linewidths similar to those of single-crystal YIG spheres. Unfortunately, at cryogenic temperatures, YIG films exhibit substantial microwave losses because the GGG substrate undergoes a phase transition into a geometrically frustrated spin-liquid state below 5 kelvin \cite{Petrenko1997Dec}. In this state, the short-range ordered spins in the GGG substrate strongly absorb external energy, a characteristic that has been exploited in commercial adiabatic demagnetization cooling applications. Consequently, in YIG/GGG structures, the presence of the spin-liquid state in GGG reduces the lifetime of spin excitations in the YIG layer, as evidenced by a larger ferromagnetic resonance (FMR) \cite{Knauer2023Apr_JAP} linewidth compared to pure YIG spheres. This presents a significant obstacle to the integration of YIG thin films in cryogenic devices. For instance, recent demonstrations of electromagnonic systems \cite{XufengPRL2014,Tabuchi2014Aug,LihuiPRL2015,Goryachev2014Nov,Huebl2013Sep,Harder_SSC_2018,Bhoi_SSP_2020,Hu2020Jan,Rameshti_PhysRep_2022,Rao2019Jul,Xu2021May,Li2019Sep,Hou2019Sep,Xu2021Mayprl}, which involve strongly coupled microwave photons and magnons and hold great promise for integration with superconducting qubits \cite{Tabuchi2015Jul,Lachance-Quirion2020Jan,Lachance-Quirion2017Jul,Wolski2020Sep,Xu2023May}, have been largely restricted to YIG spheres, while the application of YIG thin films has been limited due to their high losses at cryogenic temperatures.

\begin{figure*}[htb]
\includegraphics[width=0.65\linewidth]{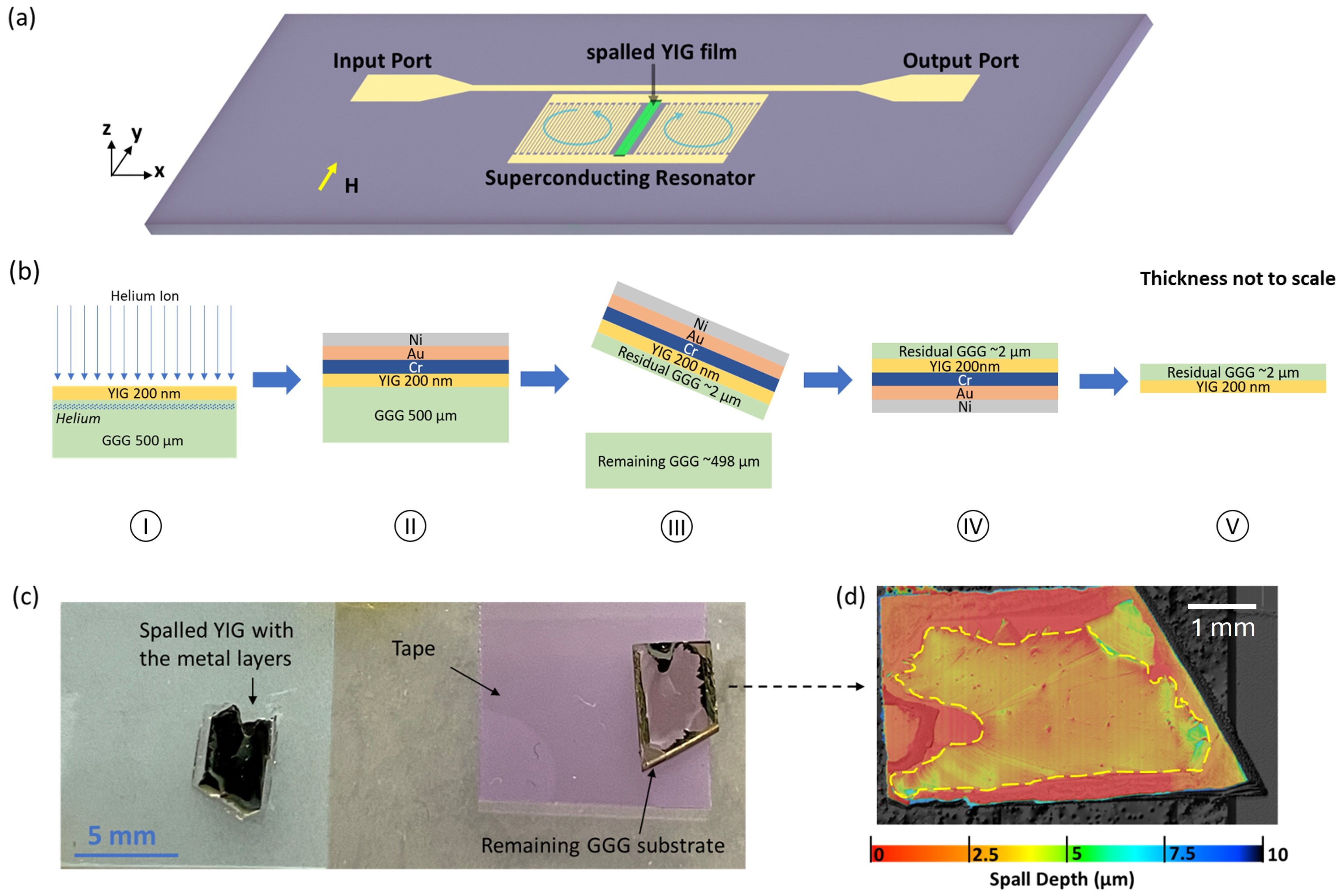}
\caption{(a) A schematic of an integrated hybrid quantum device using the spalled YIG thin film. (b) Schematics of the YIG film spalling process: (I) (Optional) helium ion implantation process with implantation depth around 7 $\mu$m in the YIG/GGG substrate.  (II) A stack of films comprising 10 nm of chromium, 70 nm of gold, and 7 $\mu$m of nickel deposited on a 200 nm YIG/500 µm GGG substrate. (III) The separation point occurs at approximately 2 µm beneath the YIG/GGG substrate's top surface, where stress accumulates, achieved by using a thermal release tape. (IV) Materials remaining on the tape, from bottom to top: 7 $\mu$m of nickel, 70 nm of gold, 10 nm of chromium, 200 nm of YIG, and 2 $\mu$m of GGG. (V) Subsequent removal of metal layers is accomplished using nickel, gold, and chromium etchants in sequence. (c) Optical image of the spalled sample after step II. Right side: spalled film adhered to the thermal release tape. Left side: the remaining GGG substrate after the spalling process. (d) Confocal microscope measurement image of the surface of the remaining GGG substrate. The outline of the spalled area is marked by the yellow dashed line.}
\label{fig1}
\end{figure*}

One promising solution to this significant challenge is to utilize YIG thin films without the GGG substrate, which can be implemented through several different technical approaches. The first method involves growing YIG thin films on alternative substrates, such as silicon \cite{Bi2013Nov,Onbasli2014Oct,Guo2022Nov,Guo2023Jun}. This approach is straightforward and could greatly facilitate device integration if successfully developed; however, the quality of YIG thin films produced in this manner is often low due to the lattice mismatch between YIG and the new substrate. The second method entails detaching epitaxially grown single-crystalline YIG from the GGG substrate \cite{Levy1998Jun,Heyroth2019Nov,Haigh2020Oct,Baity2021Jul}. While this approach yields high-quality YIG thin films, the separation processes are typically challenging and difficult to control due to the similar physical and chemical properties of YIG and GGG, which hinders broader applications. Another approach involves growing YIG structures on the GGG substrate with finite contact points, followed by a transfer process. Although high-quality YIG devices can be obtained using this method, it is challenging to produce large-area YIG thin films \cite{Heyroth2019Nov}. In this work, we present our investigation of a novel method designed to simplify the YIG detaching process, offering a new direction for the development of integrated YIG devices [Fig.\,\ref{fig1}(a)].

\section{YIG spalling and device preparation}

Our approach is based on controlled mechanical spalling \cite{Bedell2012Feb}, as illustrated by the procedures in Fig.\,\ref{fig1}(b). The substrate used is a commercially available single-crystal YIG film [200 nm thick, (111)-oriented], grown on a 500 $\mu$m-thick GGG substrate via liquid phase epitaxy (LPE). After cleaning the sample, a 10-nm-thick layer of chromium (Cr) followed by a 70-nm-thick layer of gold (Au) is deposited on the YIG surface through magnetron sputtering. Using the sputtered gold film as a seed layer, a thick film of nickel (Ni) is electroplated to a final thickness of 7 $\mu$m, utilizing the electroplating conditions outlined in Refs.\,\cite{Bedell2017Jul,Horn2024Oct}, which yields an intrinsic tensile stress of 700 MPa. This tensile stress, along with the thickness of the Ni layer, defines an equilibrium depth within the YIG/GGG substrate where steady-state crack propagation can occur \cite{Suo1989Jan}.

Using thermal release tape, the top layer stack (Ni/Au/Cr/YIG/GGG) is carefully spalled, resulting in a continuous film, as shown in Fig.\,\ref{fig1}(c), with most of the GGG substrate remaining on the bottom tape. Given that this depth exceeds the thickness of the YIG layer, the spalled YIG remains attached to a thin layer of GGG. However, by selecting YIG wafers with larger thickness (e.g., 10 $\mu$m) and optimizing the thickness and tensile stress of the Ni layer, it is possible to spall a layer of pure YIG without any residual GGG. The final metal-free device is achieved by removing all the metal layers using appropriate wet chemical etching.

We observed that under our current conditions, ion implantation into the substrate plays a crucial role in spalling. For intrinsic YIG/GGG samples, the resulting spalling depth is typically 5-10 $\mu$m. However, when the YIG/GGG sample is treated with helium ion implantation (following the conditions outlined in Refs.\,\cite{Levy1998Jun,Levy1998Oct}), thinner spalling depths are achieved along with smoother surfaces. This is demonstrated by surface profile scans of the remaining GGG substrate post-spalling [Fig.\,\ref{fig1}(d)] using a 3D laser scanning confocal microscope (Keyence VK-X1000), which reveal a spalling depth of approximately 2-3 $\mu$m. This observation can be attributed to damage within the GGG layer at the ion implantation depth (7 $\mu$m) caused by the high-energy helium ions during the implantation process, making it more prone to fracture under the elastic stress from the Ni layer and resulting in a shallower spalling depth.

\section{Room-temperature microwave characterization}

\begin{figure}[bt]
\centering
\includegraphics[width=0.98\linewidth]{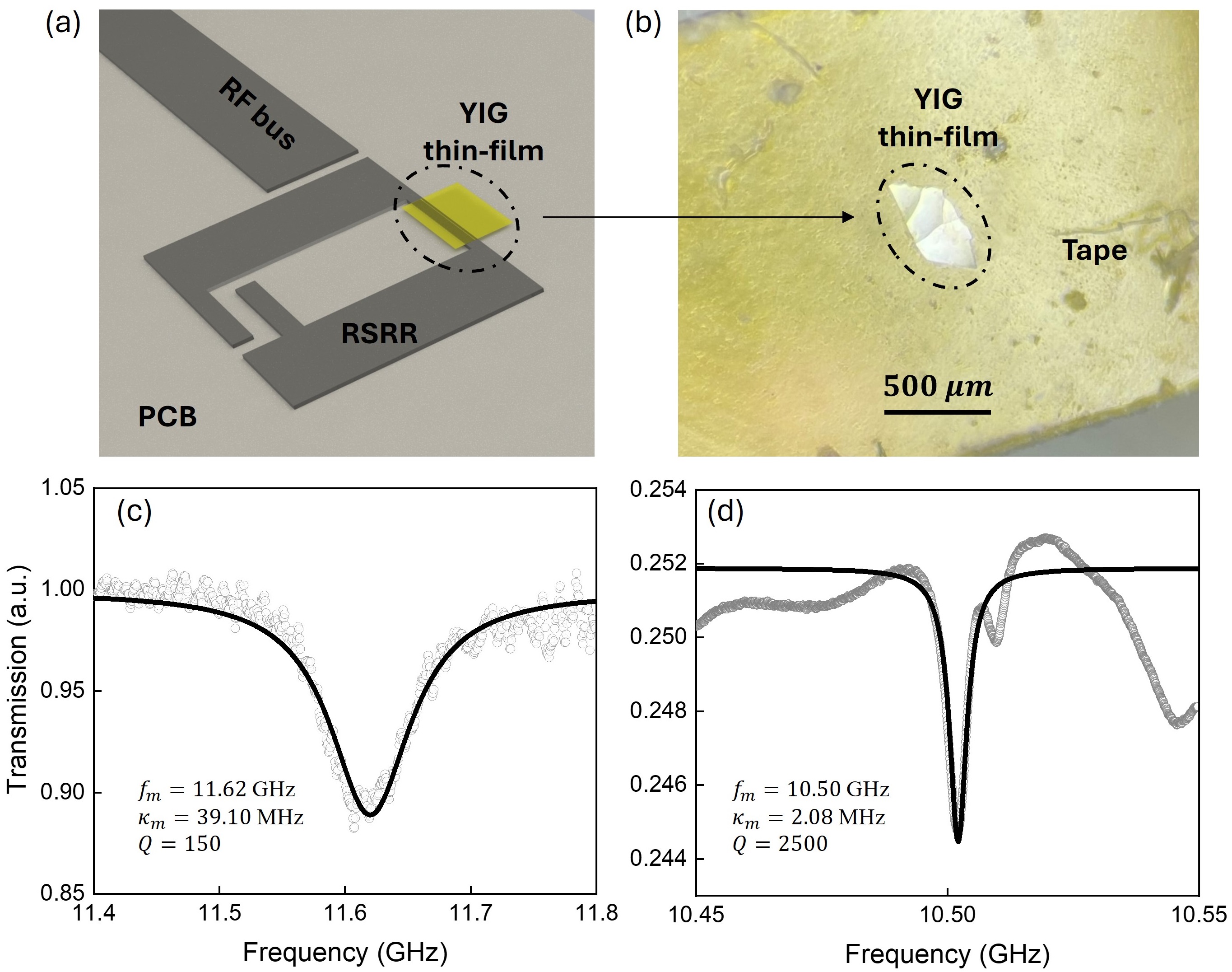}
\caption{(a) Schematics of a spalled YIG thin film bonded onto an room-temperature RSRR chip. (b) Optical image of a piece of spalled YIG (approximately $500 \mu m \times 300 \mu m$ affixed to a Kapton tape. (c) Measured reflection spectra for the RSRR device with the YIG/GGG chip attached (prior to annealing). (d) Measured reflection spectra for the RSRR device with the spalled YIG (after annealing).}
\label{fig2}
\end{figure}

To characterize the microwave performance of the spalled YIG thin film at room temperature, it is flip-bonded to a rectangular split ring resonator (RSRR) made of copper \cite{Xu2021May} for ferromagnetic resonance (FMR) measurement, as shown schematically in Fig.\,\ref{fig2}(a). The reflection spectrum of the RSRR is obtained using a vector network analyzer, and the FMR frequency of the spalled YIG is fine-tuned by adjusting the magnitude of a magnetic field applied along the out-of-plane direction. When the magnon modes approach the resonance frequency of the microwave resonator, they should become visible in the reflection spectrum.

However, when a $500\times 300 \times 2$ $\mu$m$^3$ piece of the spalled YIG film (treated with helium ion implantation) is bonded to the RSRR, no magnon modes are observed in the reflection spectra. This absence can be attributed to significant magnon losses induced by the ion implantation, which may exceed the coupling strength between the microwave resonance and the magnon mode due to the small volume of the YIG piece. To confirm this hypothesis, we tested a larger piece (approximately $5\times 5$ mm$^2$) of unspalled YIG treated with the same ion implantation process, using the same RSRR reflection measurement. With this increased YIG volume, the magnon mode is successfully detected; however, the measured data [Fig.\,\ref{fig2}(c)] shows a significantly broader linewidth (full width at half maximum, FWHM) of $\Delta \omega/2\pi =78.2$ MHz at 11.6 GHz, corresponding to a magnon dissipation rate $\kappa_m/2\pi=39.1$ MHz, which is an order of magnitude higher than previously reported values for LPE YIG thin films \cite{Dubs2017Apr_JPD,Dubs2020Feb_PRM,Pirro2014Jan_APL,Knauer2023Apr_JAP}.

The broader linewidth can be attributed to two primary dissipation sources: (1) structural damage from high-energy ion penetration into the YIG layer and (2) the accumulation of helium ions within the YIG layer. Both effects can be mitigated through an annealing process, which can restore the damaged lattice structure and expel accumulated helium ions from the YIG. We performed annealing on multiple spalled-free YIG films in ambient air, with the temperature gradually increased from room temperature to a maximum of 850 $^{\circ}$C over six hours, held for three hours, and then slowly cooled over 14 hours, allowing ample time for lattice healing and recrystallization of the YIG. Following this annealing process, the magnon mode is successfully observed in the spalled YIG device, as shown in Fig.\,\ref{fig2}(d). Numerical fitting reveals a linewidth of $\Delta\omega/2\pi=4.16$ MHz at 10.5 GHz (corresponding to $\Delta H=1.5$ Oe), comparable to those of high-quality LPE YIG thin films reported in previous studies \cite{Dubs2017Apr_JPD,Dubs2020Feb_PRM,Pirro2014Jan_APL,Knauer2023Apr_JAP}.

\begin{figure}
\includegraphics[width=0.98\linewidth]{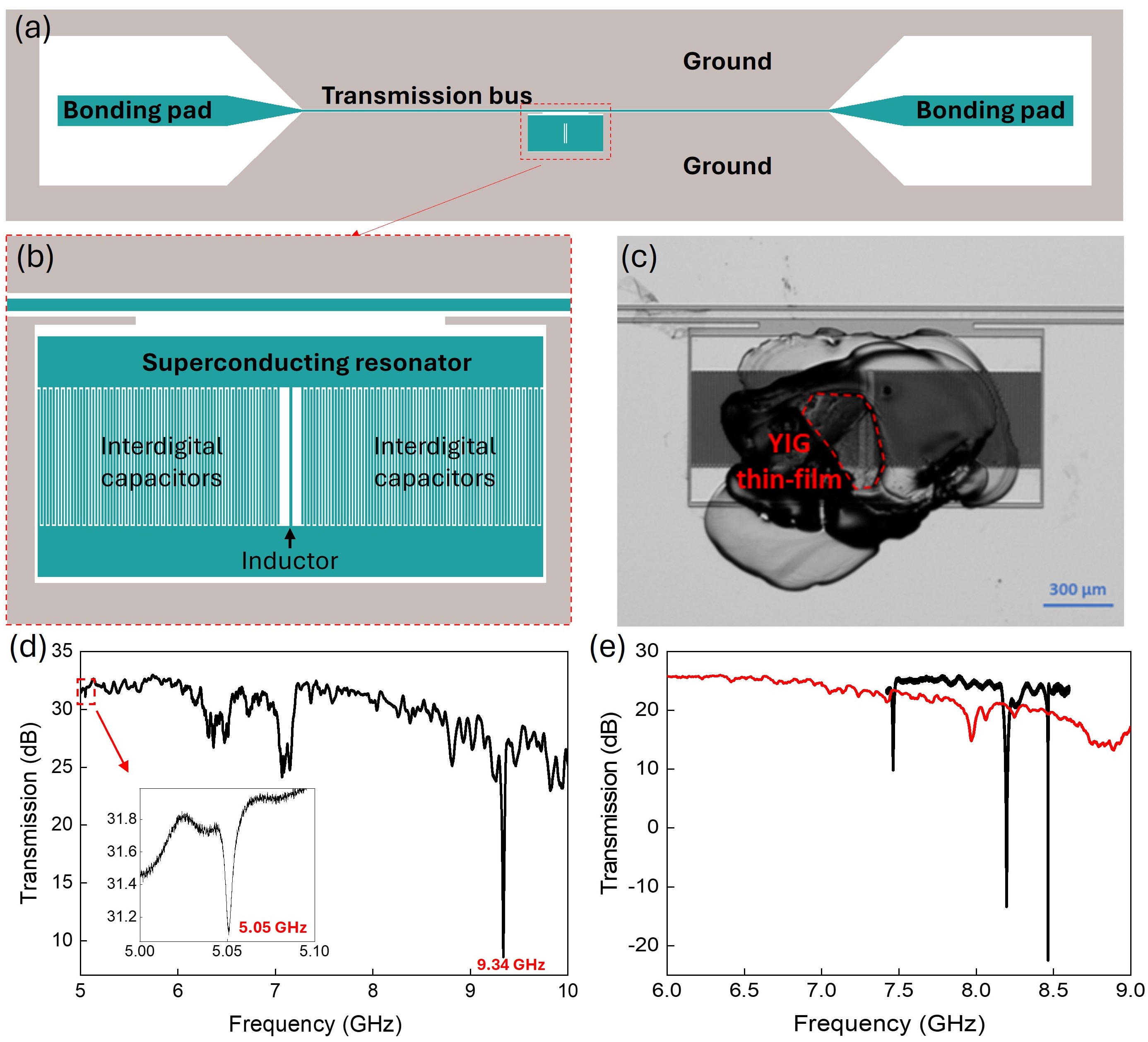}
\caption{ (a) Layout depicting the superconducting resonator coupled to a bus transmission line. (b) Zoomed-in view of the lumped element resonator design, showing the interdigital capacitors on the sides and the inductor wire in the middle. (c) Optical image displaying a superconducting resonator loaded with the spalled YIG thin film (measuring approximately 300 $\mu m \times$ 200 $\mu m$). The irregular shaded area is the GE varnish for chip bonding. (d) The measured spectrum of the chip bonded with the spalled YIG at 200 mK, revealing two microwave resonances at 5.05 GHz and 9.34 GHz, respectively. (e) The measured transmission spectrum of another superconducting resonator device at 200 mK with (red curve) and without (black curve) the unspalled YIG/GGG substrate.}
\label{fig3}
\end{figure}

\section{Low-temperature microwave characterization}
We further characterize the spalled YIG device at millikelvin (mK) temperatures to explore its potential in cryogenic quantum applications. A small piece of spalled YIG (around $300\times 200 \times 2\; \mu$m$^3$) is coupled to a superconducting microwave resonator, fabricated using a 100-nm-thick niobium layer via photolithography and dry etching. To enhance magnon-photon coupling, a lumped-element resonator with interdigital capacitors is used, inductively coupled to a bus transmission line, as shown in the layout in Figs.\,\ref{fig3}(a) and (b). The YIG flake is flip-bonded with GE varnish to cover the central inductor wire where the microwave magnetic field is strongest \cite {Hou2019Sep}, further increasing the coupling strength between the magnon and photon modes. An optical image of the assembled device is shown in Fig.\,\ref{fig3}(c), with the red dashed line indicating the shape and location of the spalled YIG film, which appears with low contrast against the yellowish GE varnish under the optical microscope.

Figure\,\ref{fig3}(d) shows the microwave transmission of the device measured at 200 millikelvin in an adiabatic demagnetization refrigerator (ADR). Two resonance modes appear as sharp dips in the transmission spectrum, similar to observations in Ref.\,\cite{Hou2019Sep}. The inset highlights the fundamental mode at 5.05 GHz, which has a lower damping rate and smaller extinction ratio. While the mode at 9.34 GHz exhibits a higher extinction ratio, it has greater dissipation and couples weakly with the YIG magnon mode, and therefore is not discussed hereafter. Importantly, the clear observation of two high-quality resonances in the YIG-loaded superconducting resonator suggests that losses from the GGG substrate have been significantly suppressed. For comparison, we measured another superconducting resonator device loaded with an unspalled YIG/GGG chip at the same temperature. Figure\,\ref{fig3}(e) clearly shows that the GGG substrate loss directly affects the superconducting microwave resonator. Prior to the YIG/GGG bonding, two resonances at 7.4 GHz and 8.5 GHz are distinctly visible [black curve in Fig.\,\ref{fig3}(e)]. However, with the addition of the 500 $\mu$m thick GGG substrate, no clear microwave resonances are observed (red curve).

\section{Strong coupling using the spalled YIG device}

\begin{figure}
\includegraphics[width=0.98\linewidth]{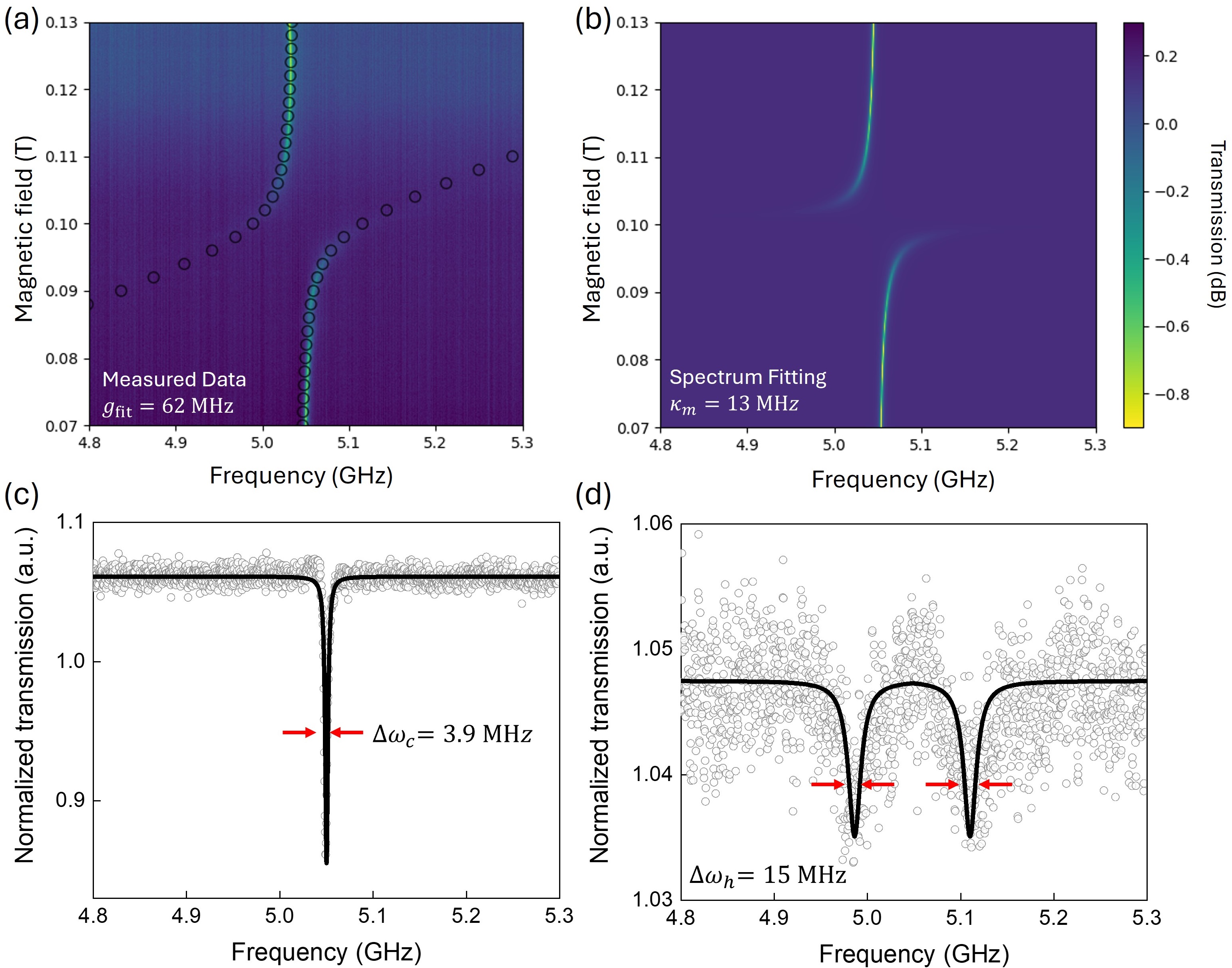}
\caption{ (a) A heatmap of the measured transmission spectrum for the superconducting resonator device shown in Fig. 3(c), showing the avoid-crossing feature. The overlaid red circles represent the calculated frequencies of magnon-photon modes. (b) A heatmap of the calculated transmission spectrum using Eq (1) with the fitted magnon linewidth. (c) Transmission data at 1300 G where the magnon mode is far detuned. The photon linewidth is extracted as 2$\kappa_c/2\pi=3.9$ MHz. (d) Transmission data at 1000 G where the magnon and the photon modes are on resonance and fully hybridized. The hybrid linewidth is fitted to be 2$\kappa_h/2\pi=15$ MHz.}
\label{fig4}
\end{figure}

To characterize the device shown in Fig.\,\ref{fig3}(c), we measured its microwave transmission while varying the strength of an applied magnetic field to tune the magnon frequency. The magnetic field is applied along the in-plane direction of the superconducting resonator chip, aligned with the orientation of the inductor wire. A clear avoided-crossing feature appears in the measured spectra, as shown in Fig.\,\ref{fig4}(a), indicating that the device has entered the strong coupling regime, where the magnon-photon coupling strength surpasses the dissipation rate of each individual mode.

Using numerical fitting based on the rotating-wave approximation (RWA) \cite{SM_Boyd2008}, the magnon-photon coupling strength $g$ is determined to be $2\pi\times 62$ MHz, with detailed procedures described in Section I of the Supplemental Material \cite{SM}. The calculated frequencies of the magnon and photon modes, shown as red dots in Fig.,\ref{fig4}(a), align well with the measured spectrum. At a magnetic field of 1300 G, where the magnon mode is far detuned from the microwave resonance, the dissipation rate of the microwave mode is extracted as $\kappa_c = \Delta \omega / 2 = 2 \pi \times 1.95$ MHz, leading to a quality factor $Q = 1295$. Since the spalled YIG film is positioned sufficiently far from the bus transmission line, its direct coupling to the line is negligible. Therefore, the magnon mode is only observed when it is close to or on resonance with the microwave mode.

The magnon dissipation rate can be estimated using the relation $\kappa_h = (\kappa_m + \kappa_c)/2$ when the modes are maximally hybridized, i.e., when the magnon and photon modes are on resonance. Here, $\kappa_h$ is the on-resonance dissipation rate of the hybrid mode, fitted from the linewidth as $\kappa_h = \Delta \omega_h / 2 = 2 \pi \times 7.50$ MHz, as shown in Fig.\,\ref{fig4}(d). Consequently, the dissipation rate of the magnon mode is determined as $\kappa_m = 2 \kappa_h - \kappa_c = 2 \pi \times 13$ MHz, corresponding to a FWHM $\Delta H=9.3$ Oe. Compared with the devices in Fig.\,\ref{fig3}(e) which consist of 200-nm-thick YIG on 500-$\mu$m-thick GGG, the magnon losses have been significantly reduced at cryogenic temperatures. Although it remains higher than the dissipation rate measured on bulk YIG samples, this may be due to the residual GGG layer attached to the YIG thin film or the rough spalled surface, which could potentially be removed through further optimization.

The extracted parameters for dissipation and coupling are consistent with the numerical fitting results [see Supplemental Material Ref.\,\cite{SM} for details] from the measured spectrum based on input-output theory \cite{Schuster2010Sep,QuantumOptics,Chen2022Dec}, as described by the equation 
\begin{equation} |S_{21}|^2 = \left| 1 - \frac{2 \kappa_\mathrm{ex}}{i \Delta_c + \kappa_c + \frac{g^2}{i \Delta_m + \kappa_m}} \right|^2, 
\label{Transmission2} 
\end{equation} 
where $g$ is the magnon-photon coupling strength, $\kappa_\mathrm{ex}$ is the external coupling rate between the microwave resonator and the bus feeding line, $\kappa_c$ ($\kappa_m$) represents the total dissipation rate of the resonator (magnon) mode, and $\Delta_c = \omega_c - \omega$ and $\Delta_m = \omega_m - \omega$ are the cavity and magnon detunings, respectively. The calculated spectra using Eq.,\ref{Transmission2} are shown in Fig.,\ref{fig4}(b), matching well with the measured result in Fig.,\ref{fig4}(a).

\section{conclusion}

In conclusion, we have demonstrated a new approach for spalling YIG thin films from GGG substrates and validated it by measuring strong magnon-photon coupling in a hybrid device. Direct comparisons with conventional YIG/GGG devices confirm the effectiveness of our approach. This work represents the first application of spalling technology to magnetic garnets, which are known for their strength and processing challenges. Compared to other methods, our spalling-based approach offers distinct advantages, including reduced material contamination and flexible thickness control. Although the linewidth of our measured device at cryogenic temperatures is still higher than that of bulk YIG, further improvement is achievable by fully removing the GGG substrate. With optimization -- such as adjusting the thickness of the initial YIG film and the stress and thickness of the Ni layer -- our method is promising for wafer-scale production of magnonic devices for quantum applications and beyond. For example, the spalled YIG films can be integrated (via bonding or pattern transfer) with on-chip microwave antennas or resonators to provide large frequency tunability even at millikelvin temperature. In addition, when combined with other novel functionalities \cite{XufengPRL2014,Tabuchi2014Aug,LihuiPRL2015,Goryachev2014Nov,Huebl2013Sep,Harder_SSC_2018,Bhoi_SSP_2020,Hu2020Jan,Rameshti_PhysRep_2022,Rao2019Jul,Xu2021May, Zhang2016Sep,Bi2011Dec,Zhu2020Oct,Seo2017Mar,An2020Feb,Zhang2016Mar,Xu2021Aug,Tabuchi2015Jul,Lachance-Quirion2020Jan,Lachance-Quirion2017Jul,Wolski2020Sep,Xu2023May, Crescini_PRL_2020,Flower_PhysDarkUniv_2019} in YIG magnonic devices, our spalled YIG thin films could enable unique properties for coupling magnons with a wide range of degrees of freedom, including optics, acoustics, and magnetics.

\section{Supplementary Material}
Further resonator simulation resutls and system modeling details can be found in the Supplementary Material.

\begin{acknowledgments}
The authors thank R. Divan, L. Stan, C. Miller, and D. Czaplewski for support in the device fabrication. X.Z. acknowledges support from NSF (2337713) and ONR Young Investigator Program (N00014-23-1-2144). Contributions by C.H. and S.G. were supported by the Vannevar Bush Fellowship received by S.G. under the program sponsored by the Office of the Undersecretary of Defense for Research and Engineering and in part by the Office of Naval Research as the Executive Manager for the grant. Work performed at the Center for Nanoscale Materials, a U.S. Department of Energy Office of Science User Facility, was supported by the U.S. DOE Office of Basic Energy Sciences, under Contract No. DE-AC02-06CH11357.
\end{acknowledgments}

\section*{AUTHOR DECLARATIONS}
\textbf{Conflict of Interest}

The authors have no conflicts to disclose.\\

\textbf{DATA AVAILABILITY}

The data that support the findings of this study are available
from the corresponding author upon reasonable request.

\vspace{12pt}


\begin{thebibliography}{56}%
\makeatletter
\providecommand \@ifxundefined [1]{%
 \@ifx{#1\undefined}
}%
\providecommand \@ifnum [1]{%
 \ifnum #1\expandafter \@firstoftwo
 \else \expandafter \@secondoftwo
 \fi
}%
\providecommand \@ifx [1]{%
 \ifx #1\expandafter \@firstoftwo
 \else \expandafter \@secondoftwo
 \fi
}%
\providecommand \natexlab [1]{#1}%
\providecommand \enquote  [1]{``#1''}%
\providecommand \bibnamefont  [1]{#1}%
\providecommand \bibfnamefont [1]{#1}%
\providecommand \citenamefont [1]{#1}%
\providecommand \href@noop [0]{\@secondoftwo}%
\providecommand \href [0]{\begingroup \@sanitize@url \@href}%
\providecommand \@href[1]{\@@startlink{#1}\@@href}%
\providecommand \@@href[1]{\endgroup#1\@@endlink}%
\providecommand \@sanitize@url [0]{\catcode `\\12\catcode `\$12\catcode `\&12\catcode `\#12\catcode `\^12\catcode `\_12\catcode `\%12\relax}%
\providecommand \@@startlink[1]{}%
\providecommand \@@endlink[0]{}%
\providecommand \url  [0]{\begingroup\@sanitize@url \@url }%
\providecommand \@url [1]{\endgroup\@href {#1}{\urlprefix }}%
\providecommand \urlprefix  [0]{URL }%
\providecommand \Eprint [0]{\href }%
\providecommand \doibase [0]{http://dx.doi.org/}%
\providecommand \selectlanguage [0]{\@gobble}%
\providecommand \bibinfo  [0]{\@secondoftwo}%
\providecommand \bibfield  [0]{\@secondoftwo}%
\providecommand \translation [1]{[#1]}%
\providecommand \BibitemOpen [0]{}%
\providecommand \bibitemStop [0]{}%
\providecommand \bibitemNoStop [0]{.\EOS\space}%
\providecommand \EOS [0]{\spacefactor3000\relax}%
\providecommand \BibitemShut  [1]{\csname bibitem#1\endcsname}%
\let\auto@bib@innerbib\@empty
\bibitem [{\citenamefont {Serga}, \citenamefont {Chumak},\ and\ \citenamefont {Hillebrands}(2010)}]{Serga2010Jun}%
  \BibitemOpen
  \bibfield  {author} {\bibinfo {author} {\bibfnamefont {A.~A.}\ \bibnamefont {Serga}}, \bibinfo {author} {\bibfnamefont {A.~V.}\ \bibnamefont {Chumak}}, \ and\ \bibinfo {author} {\bibfnamefont {B.}~\bibnamefont {Hillebrands}},\ }\bibfield  {title} {\enquote {\bibinfo {title} {{YIG magnonics}},}\ }\href {\doibase 10.1088/0022-3727/43/26/264002} {\bibfield  {journal} {\bibinfo  {journal} {J. Phys. D: Appl. Phys.}\ }\textbf {\bibinfo {volume} {43}},\ \bibinfo {pages} {264002} (\bibinfo {year} {2010})}\BibitemShut {NoStop}%
\bibitem [{\citenamefont {Awschalom}\ \emph {et~al.}(2021)\citenamefont {Awschalom}, \citenamefont {Du}, \citenamefont {He}, \citenamefont {Heremans}, \citenamefont {Hoffmann}, \citenamefont {Hou}, \citenamefont {Kurebayashi}, \citenamefont {Li}, \citenamefont {Liu}, \citenamefont {Novosad} \emph {et~al.}}]{Awschalom_IEEETransQuantEng_2021}%
  \BibitemOpen
  \bibfield  {author} {\bibinfo {author} {\bibfnamefont {D.~D.}\ \bibnamefont {Awschalom}}, \bibinfo {author} {\bibfnamefont {C.~R.}\ \bibnamefont {Du}}, \bibinfo {author} {\bibfnamefont {R.}~\bibnamefont {He}}, \bibinfo {author} {\bibfnamefont {F.~J.}\ \bibnamefont {Heremans}}, \bibinfo {author} {\bibfnamefont {A.}~\bibnamefont {Hoffmann}}, \bibinfo {author} {\bibfnamefont {J.}~\bibnamefont {Hou}}, \bibinfo {author} {\bibfnamefont {H.}~\bibnamefont {Kurebayashi}}, \bibinfo {author} {\bibfnamefont {Y.}~\bibnamefont {Li}}, \bibinfo {author} {\bibfnamefont {L.}~\bibnamefont {Liu}}, \bibinfo {author} {\bibfnamefont {V.}~\bibnamefont {Novosad}},  \emph {et~al.},\ }\bibfield  {title} {\enquote {\bibinfo {title} {Quantum engineering with hybrid magnonic systems and materials},}\ }\href@noop {} {\bibfield  {journal} {\bibinfo  {journal} {IEEE Transactions on Quantum Engineering}\ }\textbf {\bibinfo {volume} {2}},\ \bibinfo {pages} {1--36} (\bibinfo {year} {2021})}\BibitemShut {NoStop}%
\bibitem [{\citenamefont {Li}\ \emph {et~al.}(2020)\citenamefont {Li}, \citenamefont {Zhang}, \citenamefont {Tyberkevych}, \citenamefont {Kwok}, \citenamefont {Hoffmann},\ and\ \citenamefont {Novosad}}]{YiLi_JAP_2020}%
  \BibitemOpen
  \bibfield  {author} {\bibinfo {author} {\bibfnamefont {Y.}~\bibnamefont {Li}}, \bibinfo {author} {\bibfnamefont {W.}~\bibnamefont {Zhang}}, \bibinfo {author} {\bibfnamefont {V.}~\bibnamefont {Tyberkevych}}, \bibinfo {author} {\bibfnamefont {W.-K.}\ \bibnamefont {Kwok}}, \bibinfo {author} {\bibfnamefont {A.}~\bibnamefont {Hoffmann}}, \ and\ \bibinfo {author} {\bibfnamefont {V.}~\bibnamefont {Novosad}},\ }\bibfield  {title} {\enquote {\bibinfo {title} {Hybrid magnonics: Physics, circuits, and applications for coherent information processing},}\ }\href@noop {} {\bibfield  {journal} {\bibinfo  {journal} {Journal of Applied Physics}\ }\textbf {\bibinfo {volume} {128}},\ \bibinfo {pages} {130902} (\bibinfo {year} {2020})}\BibitemShut {NoStop}%
\bibitem [{\citenamefont {Zhang}(2023)}]{Zhang2023Sep}%
  \BibitemOpen
  \bibfield  {author} {\bibinfo {author} {\bibfnamefont {X.}~\bibnamefont {Zhang}},\ }\bibfield  {title} {\enquote {\bibinfo {title} {{A review of common materials for hybrid quantum magnonics}},}\ }\href {\doibase 10.1016/j.mtelec.2023.100044} {\bibfield  {journal} {\bibinfo  {journal} {Materials Today Electronics}\ }\textbf {\bibinfo {volume} {5}},\ \bibinfo {pages} {100044} (\bibinfo {year} {2023})}\BibitemShut {NoStop}%
\bibitem [{\citenamefont {Lachance-Quirion}\ \emph {et~al.}(2019)\citenamefont {Lachance-Quirion}, \citenamefont {Tabuchi}, \citenamefont {Gloppe}, \citenamefont {Usami},\ and\ \citenamefont {Nakamura}}]{Lachance_APE_2019}%
  \BibitemOpen
  \bibfield  {author} {\bibinfo {author} {\bibfnamefont {D.}~\bibnamefont {Lachance-Quirion}}, \bibinfo {author} {\bibfnamefont {Y.}~\bibnamefont {Tabuchi}}, \bibinfo {author} {\bibfnamefont {A.}~\bibnamefont {Gloppe}}, \bibinfo {author} {\bibfnamefont {K.}~\bibnamefont {Usami}}, \ and\ \bibinfo {author} {\bibfnamefont {Y.}~\bibnamefont {Nakamura}},\ }\bibfield  {title} {\enquote {\bibinfo {title} {Hybrid quantum systems based on magnonics},}\ }\href@noop {} {\bibfield  {journal} {\bibinfo  {journal} {Applied Physics Express}\ }\textbf {\bibinfo {volume} {12}},\ \bibinfo {pages} {070101} (\bibinfo {year} {2019})}\BibitemShut {NoStop}%
\bibitem [{\citenamefont {Zhang}\ \emph {et~al.}(2014)\citenamefont {Zhang}, \citenamefont {Zou}, \citenamefont {Jiang},\ and\ \citenamefont {Tang}}]{XufengPRL2014}%
  \BibitemOpen
  \bibfield  {author} {\bibinfo {author} {\bibfnamefont {X.}~\bibnamefont {Zhang}}, \bibinfo {author} {\bibfnamefont {C.-L.}\ \bibnamefont {Zou}}, \bibinfo {author} {\bibfnamefont {L.}~\bibnamefont {Jiang}}, \ and\ \bibinfo {author} {\bibfnamefont {H.~X.}\ \bibnamefont {Tang}},\ }\bibfield  {title} {\enquote {\bibinfo {title} {Strongly coupled magnons and cavity microwave photons},}\ }\href {\doibase 10.1103/PhysRevLett.113.156401} {\bibfield  {journal} {\bibinfo  {journal} {Phys. Rev. Lett.}\ }\textbf {\bibinfo {volume} {113}},\ \bibinfo {pages} {156401} (\bibinfo {year} {2014})}\BibitemShut {NoStop}%
\bibitem [{\citenamefont {Tabuchi}\ \emph {et~al.}(2014)\citenamefont {Tabuchi}, \citenamefont {Ishino}, \citenamefont {Ishikawa}, \citenamefont {Yamazaki}, \citenamefont {Usami},\ and\ \citenamefont {Nakamura}}]{Tabuchi2014Aug}%
  \BibitemOpen
  \bibfield  {author} {\bibinfo {author} {\bibfnamefont {Y.}~\bibnamefont {Tabuchi}}, \bibinfo {author} {\bibfnamefont {S.}~\bibnamefont {Ishino}}, \bibinfo {author} {\bibfnamefont {T.}~\bibnamefont {Ishikawa}}, \bibinfo {author} {\bibfnamefont {R.}~\bibnamefont {Yamazaki}}, \bibinfo {author} {\bibfnamefont {K.}~\bibnamefont {Usami}}, \ and\ \bibinfo {author} {\bibfnamefont {Y.}~\bibnamefont {Nakamura}},\ }\bibfield  {title} {\enquote {\bibinfo {title} {{Hybridizing Ferromagnetic Magnons and Microwave Photons in the Quantum Limit}},}\ }\href {\doibase 10.1103/PhysRevLett.113.083603} {\bibfield  {journal} {\bibinfo  {journal} {Phys. Rev. Lett.}\ }\textbf {\bibinfo {volume} {113}},\ \bibinfo {pages} {083603} (\bibinfo {year} {2014})}\BibitemShut {NoStop}%
\bibitem [{\citenamefont {Bai}\ \emph {et~al.}(2015)\citenamefont {Bai}, \citenamefont {Harder}, \citenamefont {Chen}, \citenamefont {Fan}, \citenamefont {Xiao},\ and\ \citenamefont {Hu}}]{LihuiPRL2015}%
  \BibitemOpen
  \bibfield  {author} {\bibinfo {author} {\bibfnamefont {L.}~\bibnamefont {Bai}}, \bibinfo {author} {\bibfnamefont {M.}~\bibnamefont {Harder}}, \bibinfo {author} {\bibfnamefont {Y.~P.}\ \bibnamefont {Chen}}, \bibinfo {author} {\bibfnamefont {X.}~\bibnamefont {Fan}}, \bibinfo {author} {\bibfnamefont {J.~Q.}\ \bibnamefont {Xiao}}, \ and\ \bibinfo {author} {\bibfnamefont {C.-M.}\ \bibnamefont {Hu}},\ }\bibfield  {title} {\enquote {\bibinfo {title} {Spin pumping in electrodynamically coupled magnon-photon systems},}\ }\href {\doibase 10.1103/PhysRevLett.114.227201} {\bibfield  {journal} {\bibinfo  {journal} {Phys. Rev. Lett.}\ }\textbf {\bibinfo {volume} {114}},\ \bibinfo {pages} {227201} (\bibinfo {year} {2015})}\BibitemShut {NoStop}%
\bibitem [{\citenamefont {Goryachev}\ \emph {et~al.}(2014)\citenamefont {Goryachev}, \citenamefont {Farr}, \citenamefont {Creedon}, \citenamefont {Fan}, \citenamefont {Kostylev},\ and\ \citenamefont {Tobar}}]{Goryachev2014Nov}%
  \BibitemOpen
  \bibfield  {author} {\bibinfo {author} {\bibfnamefont {M.}~\bibnamefont {Goryachev}}, \bibinfo {author} {\bibfnamefont {W.~G.}\ \bibnamefont {Farr}}, \bibinfo {author} {\bibfnamefont {D.~L.}\ \bibnamefont {Creedon}}, \bibinfo {author} {\bibfnamefont {Y.}~\bibnamefont {Fan}}, \bibinfo {author} {\bibfnamefont {M.}~\bibnamefont {Kostylev}}, \ and\ \bibinfo {author} {\bibfnamefont {M.~E.}\ \bibnamefont {Tobar}},\ }\bibfield  {title} {\enquote {\bibinfo {title} {{High-Cooperativity Cavity QED with Magnons at Microwave Frequencies}},}\ }\href {\doibase 10.1103/PhysRevApplied.2.054002} {\bibfield  {journal} {\bibinfo  {journal} {Phys. Rev. Appl.}\ }\textbf {\bibinfo {volume} {2}},\ \bibinfo {pages} {054002} (\bibinfo {year} {2014})}\BibitemShut {NoStop}%
\bibitem [{\citenamefont {Huebl}\ \emph {et~al.}(2013)\citenamefont {Huebl}, \citenamefont {Zollitsch}, \citenamefont {Lotze}, \citenamefont {Hocke}, \citenamefont {Greifenstein}, \citenamefont {Marx}, \citenamefont {Gross},\ and\ \citenamefont {Goennenwein}}]{Huebl2013Sep}%
  \BibitemOpen
  \bibfield  {author} {\bibinfo {author} {\bibfnamefont {H.}~\bibnamefont {Huebl}}, \bibinfo {author} {\bibfnamefont {C.~W.}\ \bibnamefont {Zollitsch}}, \bibinfo {author} {\bibfnamefont {J.}~\bibnamefont {Lotze}}, \bibinfo {author} {\bibfnamefont {F.}~\bibnamefont {Hocke}}, \bibinfo {author} {\bibfnamefont {M.}~\bibnamefont {Greifenstein}}, \bibinfo {author} {\bibfnamefont {A.}~\bibnamefont {Marx}}, \bibinfo {author} {\bibfnamefont {R.}~\bibnamefont {Gross}}, \ and\ \bibinfo {author} {\bibfnamefont {S.~T.~B.}\ \bibnamefont {Goennenwein}},\ }\bibfield  {title} {\enquote {\bibinfo {title} {{High Cooperativity in Coupled Microwave Resonator Ferrimagnetic Insulator Hybrids}},}\ }\href {\doibase 10.1103/PhysRevLett.111.127003} {\bibfield  {journal} {\bibinfo  {journal} {Phys. Rev. Lett.}\ }\textbf {\bibinfo {volume} {111}},\ \bibinfo {pages} {127003} (\bibinfo {year} {2013})}\BibitemShut {NoStop}%
\bibitem [{\citenamefont {Harder}\ and\ \citenamefont {Hu}(2018)}]{Harder_SSC_2018}%
  \BibitemOpen
  \bibfield  {author} {\bibinfo {author} {\bibfnamefont {M.}~\bibnamefont {Harder}}\ and\ \bibinfo {author} {\bibfnamefont {C.-M.}\ \bibnamefont {Hu}},\ }\bibfield  {title} {\enquote {\bibinfo {title} {Cavity spintronics: an early review of recent progress in the study of magnon--photon level repulsion},}\ }\href@noop {} {\bibfield  {journal} {\bibinfo  {journal} {Solid State Physics}\ }\textbf {\bibinfo {volume} {69}},\ \bibinfo {pages} {47--121} (\bibinfo {year} {2018})}\BibitemShut {NoStop}%
\bibitem [{\citenamefont {Bhoi}\ and\ \citenamefont {Kim}(2020)}]{Bhoi_SSP_2020}%
  \BibitemOpen
  \bibfield  {author} {\bibinfo {author} {\bibfnamefont {B.}~\bibnamefont {Bhoi}}\ and\ \bibinfo {author} {\bibfnamefont {S.-K.}\ \bibnamefont {Kim}},\ }\bibfield  {title} {\enquote {\bibinfo {title} {Roadmap for photon-magnon coupling and its applications},}\ }in\ \href@noop {} {\emph {\bibinfo {booktitle} {Solid State Physics}}},\ Vol.~\bibinfo {volume} {71}\ (\bibinfo  {publisher} {Elsevier},\ \bibinfo {year} {2020})\ pp.\ \bibinfo {pages} {39--71}\BibitemShut {NoStop}%
\bibitem [{\citenamefont {Hu}(2020)}]{Hu2020Jan}%
  \BibitemOpen
  \bibfield  {author} {\bibinfo {author} {\bibfnamefont {C.-M.}\ \bibnamefont {Hu}},\ }\bibfield  {title} {\enquote {\bibinfo {title} {{The 2020 roadmap for spin cavitronics}},}\ }in\ \href {\doibase 10.1016/bs.ssp.2020.09.003} {\emph {\bibinfo {booktitle} {{Solid State Physics}}}},\ Vol.~\bibinfo {volume} {71}\ (\bibinfo  {publisher} {Academic Press},\ \bibinfo {address} {Cambridge, MA, USA},\ \bibinfo {year} {2020})\ pp.\ \bibinfo {pages} {117--121}\BibitemShut {NoStop}%
\bibitem [{\citenamefont {Rameshti}\ \emph {et~al.}(2022)\citenamefont {Rameshti}, \citenamefont {Kusminskiy}, \citenamefont {Haigh}, \citenamefont {Usami}, \citenamefont {Lachance-Quirion}, \citenamefont {Nakamura}, \citenamefont {Hu}, \citenamefont {Tang}, \citenamefont {Bauer},\ and\ \citenamefont {Blanter}}]{Rameshti_PhysRep_2022}%
  \BibitemOpen
  \bibfield  {author} {\bibinfo {author} {\bibfnamefont {B.~Z.}\ \bibnamefont {Rameshti}}, \bibinfo {author} {\bibfnamefont {S.~V.}\ \bibnamefont {Kusminskiy}}, \bibinfo {author} {\bibfnamefont {J.~A.}\ \bibnamefont {Haigh}}, \bibinfo {author} {\bibfnamefont {K.}~\bibnamefont {Usami}}, \bibinfo {author} {\bibfnamefont {D.}~\bibnamefont {Lachance-Quirion}}, \bibinfo {author} {\bibfnamefont {Y.}~\bibnamefont {Nakamura}}, \bibinfo {author} {\bibfnamefont {C.-M.}\ \bibnamefont {Hu}}, \bibinfo {author} {\bibfnamefont {H.~X.}\ \bibnamefont {Tang}}, \bibinfo {author} {\bibfnamefont {G.~E.}\ \bibnamefont {Bauer}}, \ and\ \bibinfo {author} {\bibfnamefont {Y.~M.}\ \bibnamefont {Blanter}},\ }\bibfield  {title} {\enquote {\bibinfo {title} {Cavity magnonics},}\ }\href@noop {} {\bibfield  {journal} {\bibinfo  {journal} {Physics Reports}\ }\textbf {\bibinfo {volume} {979}},\ \bibinfo {pages} {1--61} (\bibinfo {year} {2022})}\BibitemShut {NoStop}%
\bibitem [{\citenamefont {Rao}\ \emph {et~al.}(2019)\citenamefont {Rao}, \citenamefont {Kaur}, \citenamefont {Yao}, \citenamefont {Edwards}, \citenamefont {Zhao}, \citenamefont {Fan}, \citenamefont {Xue}, \citenamefont {Silva}, \citenamefont {Gui},\ and\ \citenamefont {Hu}}]{Rao2019Jul}%
  \BibitemOpen
  \bibfield  {author} {\bibinfo {author} {\bibfnamefont {J.~W.}\ \bibnamefont {Rao}}, \bibinfo {author} {\bibfnamefont {S.}~\bibnamefont {Kaur}}, \bibinfo {author} {\bibfnamefont {B.~M.}\ \bibnamefont {Yao}}, \bibinfo {author} {\bibfnamefont {E.~R.~J.}\ \bibnamefont {Edwards}}, \bibinfo {author} {\bibfnamefont {Y.~T.}\ \bibnamefont {Zhao}}, \bibinfo {author} {\bibfnamefont {X.}~\bibnamefont {Fan}}, \bibinfo {author} {\bibfnamefont {D.}~\bibnamefont {Xue}}, \bibinfo {author} {\bibfnamefont {T.~J.}\ \bibnamefont {Silva}}, \bibinfo {author} {\bibfnamefont {Y.~S.}\ \bibnamefont {Gui}}, \ and\ \bibinfo {author} {\bibfnamefont {C.-M.}\ \bibnamefont {Hu}},\ }\bibfield  {title} {\enquote {\bibinfo {title} {{Analogue of dynamic Hall effect in cavity magnon polariton system and coherently controlled logic device}},}\ }\href {\doibase 10.1038/s41467-019-11021-2} {\bibfield  {journal} {\bibinfo  {journal} {Nat. Commun.}\ }\textbf {\bibinfo {volume} {10}},\ \bibinfo {pages} {1--7} (\bibinfo {year} {2019})}\BibitemShut
  {NoStop}%
\bibitem [{\citenamefont {Xu}\ \emph {et~al.}(2021{\natexlab{a}})\citenamefont {Xu}, \citenamefont {Zhong}, \citenamefont {Han}, \citenamefont {Jin}, \citenamefont {Jiang},\ and\ \citenamefont {Zhang}}]{Xu2021May}%
  \BibitemOpen
  \bibfield  {author} {\bibinfo {author} {\bibfnamefont {J.}~\bibnamefont {Xu}}, \bibinfo {author} {\bibfnamefont {C.}~\bibnamefont {Zhong}}, \bibinfo {author} {\bibfnamefont {X.}~\bibnamefont {Han}}, \bibinfo {author} {\bibfnamefont {D.}~\bibnamefont {Jin}}, \bibinfo {author} {\bibfnamefont {L.}~\bibnamefont {Jiang}}, \ and\ \bibinfo {author} {\bibfnamefont {X.}~\bibnamefont {Zhang}},\ }\bibfield  {title} {\enquote {\bibinfo {title} {{Coherent Gate Operations in Hybrid Magnonics}},}\ }\href {\doibase 10.1103/PhysRevLett.126.207202} {\bibfield  {journal} {\bibinfo  {journal} {Phys. Rev. Lett.}\ }\textbf {\bibinfo {volume} {126}},\ \bibinfo {pages} {207202} (\bibinfo {year} {2021}{\natexlab{a}})}\BibitemShut {NoStop}%
\bibitem [{\citenamefont {Zhang}\ \emph {et~al.}(2016{\natexlab{a}})\citenamefont {Zhang}, \citenamefont {Zhu}, \citenamefont {Zou},\ and\ \citenamefont {Tang}}]{Zhang2016Sep}%
  \BibitemOpen
  \bibfield  {author} {\bibinfo {author} {\bibfnamefont {X.}~\bibnamefont {Zhang}}, \bibinfo {author} {\bibfnamefont {N.}~\bibnamefont {Zhu}}, \bibinfo {author} {\bibfnamefont {C.-L.}\ \bibnamefont {Zou}}, \ and\ \bibinfo {author} {\bibfnamefont {H.~X.}\ \bibnamefont {Tang}},\ }\bibfield  {title} {\enquote {\bibinfo {title} {{Optomagnonic Whispering Gallery Microresonators}},}\ }\href {\doibase 10.1103/PhysRevLett.117.123605} {\bibfield  {journal} {\bibinfo  {journal} {Phys. Rev. Lett.}\ }\textbf {\bibinfo {volume} {117}},\ \bibinfo {pages} {123605} (\bibinfo {year} {2016}{\natexlab{a}})}\BibitemShut {NoStop}%
\bibitem [{\citenamefont {Bi}\ \emph {et~al.}(2011)\citenamefont {Bi}, \citenamefont {Hu}, \citenamefont {Jiang}, \citenamefont {Kim}, \citenamefont {Dionne}, \citenamefont {Kimerling},\ and\ \citenamefont {Ross}}]{Bi2011Dec}%
  \BibitemOpen
  \bibfield  {author} {\bibinfo {author} {\bibfnamefont {L.}~\bibnamefont {Bi}}, \bibinfo {author} {\bibfnamefont {J.}~\bibnamefont {Hu}}, \bibinfo {author} {\bibfnamefont {P.}~\bibnamefont {Jiang}}, \bibinfo {author} {\bibfnamefont {D.~H.}\ \bibnamefont {Kim}}, \bibinfo {author} {\bibfnamefont {G.~F.}\ \bibnamefont {Dionne}}, \bibinfo {author} {\bibfnamefont {L.~C.}\ \bibnamefont {Kimerling}}, \ and\ \bibinfo {author} {\bibfnamefont {C.~A.}\ \bibnamefont {Ross}},\ }\bibfield  {title} {\enquote {\bibinfo {title} {{On-chip optical isolation in monolithically integrated non-reciprocal optical resonators}},}\ }\href {\doibase 10.1038/nphoton.2011.270} {\bibfield  {journal} {\bibinfo  {journal} {Nat. Photonics}\ }\textbf {\bibinfo {volume} {5}},\ \bibinfo {pages} {758--762} (\bibinfo {year} {2011})}\BibitemShut {NoStop}%
\bibitem [{\citenamefont {Zhu}\ \emph {et~al.}(2020)\citenamefont {Zhu}, \citenamefont {Zhang}, \citenamefont {Zhang}, \citenamefont {Han}, \citenamefont {Han}, \citenamefont {Zou}, \citenamefont {Zou}, \citenamefont {Zhong}, \citenamefont {Zhong}, \citenamefont {Wang}, \citenamefont {Wang}, \citenamefont {Jiang}, \citenamefont {Jiang},\ and\ \citenamefont {Tang}}]{Zhu2020Oct}%
  \BibitemOpen
  \bibfield  {author} {\bibinfo {author} {\bibfnamefont {N.}~\bibnamefont {Zhu}}, \bibinfo {author} {\bibfnamefont {X.}~\bibnamefont {Zhang}}, \bibinfo {author} {\bibfnamefont {X.}~\bibnamefont {Zhang}}, \bibinfo {author} {\bibfnamefont {X.}~\bibnamefont {Han}}, \bibinfo {author} {\bibfnamefont {X.}~\bibnamefont {Han}}, \bibinfo {author} {\bibfnamefont {C.-L.}\ \bibnamefont {Zou}}, \bibinfo {author} {\bibfnamefont {C.-L.}\ \bibnamefont {Zou}}, \bibinfo {author} {\bibfnamefont {C.}~\bibnamefont {Zhong}}, \bibinfo {author} {\bibfnamefont {C.}~\bibnamefont {Zhong}}, \bibinfo {author} {\bibfnamefont {C.-H.}\ \bibnamefont {Wang}}, \bibinfo {author} {\bibfnamefont {C.-H.}\ \bibnamefont {Wang}}, \bibinfo {author} {\bibfnamefont {L.}~\bibnamefont {Jiang}}, \bibinfo {author} {\bibfnamefont {L.}~\bibnamefont {Jiang}}, \ and\ \bibinfo {author} {\bibfnamefont {H.~X.}\ \bibnamefont {Tang}},\ }\bibfield  {title} {\enquote {\bibinfo {title} {{Waveguide cavity optomagnonics for microwave-to-optics conversion}},}\ }\href
  {\doibase 10.1364/OPTICA.397967} {\bibfield  {journal} {\bibinfo  {journal} {Optica}\ }\textbf {\bibinfo {volume} {7}},\ \bibinfo {pages} {1291--1297} (\bibinfo {year} {2020})}\BibitemShut {NoStop}%
\bibitem [{\citenamefont {Seo}\ \emph {et~al.}(2017)\citenamefont {Seo}, \citenamefont {Harii}, \citenamefont {Takahashi}, \citenamefont {Chudo}, \citenamefont {Oyanagi}, \citenamefont {Qiu}, \citenamefont {Ono}, \citenamefont {Shiomi},\ and\ \citenamefont {Saitoh}}]{Seo2017Mar}%
  \BibitemOpen
  \bibfield  {author} {\bibinfo {author} {\bibfnamefont {Y.-J.}\ \bibnamefont {Seo}}, \bibinfo {author} {\bibfnamefont {K.}~\bibnamefont {Harii}}, \bibinfo {author} {\bibfnamefont {R.}~\bibnamefont {Takahashi}}, \bibinfo {author} {\bibfnamefont {H.}~\bibnamefont {Chudo}}, \bibinfo {author} {\bibfnamefont {K.}~\bibnamefont {Oyanagi}}, \bibinfo {author} {\bibfnamefont {Z.}~\bibnamefont {Qiu}}, \bibinfo {author} {\bibfnamefont {T.}~\bibnamefont {Ono}}, \bibinfo {author} {\bibfnamefont {Y.}~\bibnamefont {Shiomi}}, \ and\ \bibinfo {author} {\bibfnamefont {E.}~\bibnamefont {Saitoh}},\ }\bibfield  {title} {\enquote {\bibinfo {title} {{Fabrication and magnetic control of Y3Fe5O12 cantilevers}},}\ }\href {\doibase 10.1063/1.4979553} {\bibfield  {journal} {\bibinfo  {journal} {Appl. Phys. Lett.}\ }\textbf {\bibinfo {volume} {110}} (\bibinfo {year} {2017}),\ 10.1063/1.4979553}\BibitemShut {NoStop}%
\bibitem [{\citenamefont {An}\ \emph {et~al.}(2020)\citenamefont {An}, \citenamefont {Litvinenko}, \citenamefont {Kohno}, \citenamefont {Fuad}, \citenamefont {Naletov}, \citenamefont {Vila}, \citenamefont {Ebels}, \citenamefont {de~Loubens}, \citenamefont {Hurdequint}, \citenamefont {Beaulieu}, \citenamefont {Ben~Youssef}, \citenamefont {Vukadinovic}, \citenamefont {Bauer}, \citenamefont {Slavin}, \citenamefont {Tiberkevich},\ and\ \citenamefont {Klein}}]{An2020Feb}%
  \BibitemOpen
  \bibfield  {author} {\bibinfo {author} {\bibfnamefont {K.}~\bibnamefont {An}}, \bibinfo {author} {\bibfnamefont {A.~N.}\ \bibnamefont {Litvinenko}}, \bibinfo {author} {\bibfnamefont {R.}~\bibnamefont {Kohno}}, \bibinfo {author} {\bibfnamefont {A.~A.}\ \bibnamefont {Fuad}}, \bibinfo {author} {\bibfnamefont {V.~V.}\ \bibnamefont {Naletov}}, \bibinfo {author} {\bibfnamefont {L.}~\bibnamefont {Vila}}, \bibinfo {author} {\bibfnamefont {U.}~\bibnamefont {Ebels}}, \bibinfo {author} {\bibfnamefont {G.}~\bibnamefont {de~Loubens}}, \bibinfo {author} {\bibfnamefont {H.}~\bibnamefont {Hurdequint}}, \bibinfo {author} {\bibfnamefont {N.}~\bibnamefont {Beaulieu}}, \bibinfo {author} {\bibfnamefont {J.}~\bibnamefont {Ben~Youssef}}, \bibinfo {author} {\bibfnamefont {N.}~\bibnamefont {Vukadinovic}}, \bibinfo {author} {\bibfnamefont {G.~E.~W.}\ \bibnamefont {Bauer}}, \bibinfo {author} {\bibfnamefont {A.~N.}\ \bibnamefont {Slavin}}, \bibinfo {author} {\bibfnamefont {V.~S.}\ \bibnamefont {Tiberkevich}}, \ and\ \bibinfo {author}
  {\bibfnamefont {O.}~\bibnamefont {Klein}},\ }\bibfield  {title} {\enquote {\bibinfo {title} {{Coherent long-range transfer of angular momentum between magnon Kittel modes by phonons}},}\ }\href {\doibase 10.1103/PhysRevB.101.060407} {\bibfield  {journal} {\bibinfo  {journal} {Phys. Rev. B}\ }\textbf {\bibinfo {volume} {101}},\ \bibinfo {pages} {060407} (\bibinfo {year} {2020})}\BibitemShut {NoStop}%
\bibitem [{\citenamefont {Zhang}\ \emph {et~al.}(2016{\natexlab{b}})\citenamefont {Zhang}, \citenamefont {Zou}, \citenamefont {Jiang},\ and\ \citenamefont {Tang}}]{Zhang2016Mar}%
  \BibitemOpen
  \bibfield  {author} {\bibinfo {author} {\bibfnamefont {X.}~\bibnamefont {Zhang}}, \bibinfo {author} {\bibfnamefont {C.-L.}\ \bibnamefont {Zou}}, \bibinfo {author} {\bibfnamefont {L.}~\bibnamefont {Jiang}}, \ and\ \bibinfo {author} {\bibfnamefont {H.~X.}\ \bibnamefont {Tang}},\ }\bibfield  {title} {\enquote {\bibinfo {title} {{Cavity magnomechanics}},}\ }\href {\doibase 10.1126/sciadv.1501286} {\bibfield  {journal} {\bibinfo  {journal} {Sci. Adv.}\ }\textbf {\bibinfo {volume} {2}} (\bibinfo {year} {2016}{\natexlab{b}}),\ 10.1126/sciadv.1501286}\BibitemShut {NoStop}%
\bibitem [{\citenamefont {Xu}\ \emph {et~al.}(2021{\natexlab{b}})\citenamefont {Xu}, \citenamefont {Zhong}, \citenamefont {Zhou}, \citenamefont {Han}, \citenamefont {Jin}, \citenamefont {Gray}, \citenamefont {Jiang},\ and\ \citenamefont {Zhang}}]{Xu2021Aug}%
  \BibitemOpen
  \bibfield  {author} {\bibinfo {author} {\bibfnamefont {J.}~\bibnamefont {Xu}}, \bibinfo {author} {\bibfnamefont {C.}~\bibnamefont {Zhong}}, \bibinfo {author} {\bibfnamefont {X.}~\bibnamefont {Zhou}}, \bibinfo {author} {\bibfnamefont {X.}~\bibnamefont {Han}}, \bibinfo {author} {\bibfnamefont {D.}~\bibnamefont {Jin}}, \bibinfo {author} {\bibfnamefont {S.~K.}\ \bibnamefont {Gray}}, \bibinfo {author} {\bibfnamefont {L.}~\bibnamefont {Jiang}}, \ and\ \bibinfo {author} {\bibfnamefont {X.}~\bibnamefont {Zhang}},\ }\bibfield  {title} {\enquote {\bibinfo {title} {{Coherent Pulse Echo in Hybrid Magnonics with Multimode Phonons}},}\ }\href {\doibase 10.1103/PhysRevApplied.16.024009} {\bibfield  {journal} {\bibinfo  {journal} {Phys. Rev. Appl.}\ }\textbf {\bibinfo {volume} {16}},\ \bibinfo {pages} {024009} (\bibinfo {year} {2021}{\natexlab{b}})}\BibitemShut {NoStop}%
\bibitem [{\citenamefont {Tabuchi}\ \emph {et~al.}(2015)\citenamefont {Tabuchi}, \citenamefont {Ishino}, \citenamefont {Noguchi}, \citenamefont {Ishikawa}, \citenamefont {Yamazaki}, \citenamefont {Usami},\ and\ \citenamefont {Nakamura}}]{Tabuchi2015Jul}%
  \BibitemOpen
  \bibfield  {author} {\bibinfo {author} {\bibfnamefont {Y.}~\bibnamefont {Tabuchi}}, \bibinfo {author} {\bibfnamefont {S.}~\bibnamefont {Ishino}}, \bibinfo {author} {\bibfnamefont {A.}~\bibnamefont {Noguchi}}, \bibinfo {author} {\bibfnamefont {T.}~\bibnamefont {Ishikawa}}, \bibinfo {author} {\bibfnamefont {R.}~\bibnamefont {Yamazaki}}, \bibinfo {author} {\bibfnamefont {K.}~\bibnamefont {Usami}}, \ and\ \bibinfo {author} {\bibfnamefont {Y.}~\bibnamefont {Nakamura}},\ }\bibfield  {title} {\enquote {\bibinfo {title} {{Coherent coupling between a ferromagnetic magnon and a superconducting qubit}},}\ }\href {\doibase 10.1126/science.aaa3693} {\bibfield  {journal} {\bibinfo  {journal} {Science}\ }\textbf {\bibinfo {volume} {349}},\ \bibinfo {pages} {405--408} (\bibinfo {year} {2015})}\BibitemShut {NoStop}%
\bibitem [{\citenamefont {Lachance-Quirion}\ \emph {et~al.}(2020)\citenamefont {Lachance-Quirion}, \citenamefont {Wolski}, \citenamefont {Tabuchi}, \citenamefont {Kono}, \citenamefont {Usami},\ and\ \citenamefont {Nakamura}}]{Lachance-Quirion2020Jan}%
  \BibitemOpen
  \bibfield  {author} {\bibinfo {author} {\bibfnamefont {D.}~\bibnamefont {Lachance-Quirion}}, \bibinfo {author} {\bibfnamefont {S.~P.}\ \bibnamefont {Wolski}}, \bibinfo {author} {\bibfnamefont {Y.}~\bibnamefont {Tabuchi}}, \bibinfo {author} {\bibfnamefont {S.}~\bibnamefont {Kono}}, \bibinfo {author} {\bibfnamefont {K.}~\bibnamefont {Usami}}, \ and\ \bibinfo {author} {\bibfnamefont {Y.}~\bibnamefont {Nakamura}},\ }\bibfield  {title} {\enquote {\bibinfo {title} {{Entanglement-based single-shot detection of a single magnon with a superconducting qubit}},}\ }\href {\doibase 10.1126/science.aaz9236} {\bibfield  {journal} {\bibinfo  {journal} {Science}\ }\textbf {\bibinfo {volume} {367}},\ \bibinfo {pages} {425--428} (\bibinfo {year} {2020})}\BibitemShut {NoStop}%
\bibitem [{\citenamefont {Lachance-Quirion}\ \emph {et~al.}(2017)\citenamefont {Lachance-Quirion}, \citenamefont {Tabuchi}, \citenamefont {Ishino}, \citenamefont {Noguchi}, \citenamefont {Ishikawa}, \citenamefont {Yamazaki},\ and\ \citenamefont {Nakamura}}]{Lachance-Quirion2017Jul}%
  \BibitemOpen
  \bibfield  {author} {\bibinfo {author} {\bibfnamefont {D.}~\bibnamefont {Lachance-Quirion}}, \bibinfo {author} {\bibfnamefont {Y.}~\bibnamefont {Tabuchi}}, \bibinfo {author} {\bibfnamefont {S.}~\bibnamefont {Ishino}}, \bibinfo {author} {\bibfnamefont {A.}~\bibnamefont {Noguchi}}, \bibinfo {author} {\bibfnamefont {T.}~\bibnamefont {Ishikawa}}, \bibinfo {author} {\bibfnamefont {R.}~\bibnamefont {Yamazaki}}, \ and\ \bibinfo {author} {\bibfnamefont {Y.}~\bibnamefont {Nakamura}},\ }\bibfield  {title} {\enquote {\bibinfo {title} {{Resolving quanta of collective spin excitations in a millimeter-sized ferromagnet}},}\ }\href {\doibase 10.1126/sciadv.1603150} {\bibfield  {journal} {\bibinfo  {journal} {Sci. Adv.}\ }\textbf {\bibinfo {volume} {3}} (\bibinfo {year} {2017}),\ 10.1126/sciadv.1603150}\BibitemShut {NoStop}%
\bibitem [{\citenamefont {Wolski}\ \emph {et~al.}(2020)\citenamefont {Wolski}, \citenamefont {Lachance-Quirion}, \citenamefont {Tabuchi}, \citenamefont {Kono}, \citenamefont {Noguchi}, \citenamefont {Usami},\ and\ \citenamefont {Nakamura}}]{Wolski2020Sep}%
  \BibitemOpen
  \bibfield  {author} {\bibinfo {author} {\bibfnamefont {S.~P.}\ \bibnamefont {Wolski}}, \bibinfo {author} {\bibfnamefont {D.}~\bibnamefont {Lachance-Quirion}}, \bibinfo {author} {\bibfnamefont {Y.}~\bibnamefont {Tabuchi}}, \bibinfo {author} {\bibfnamefont {S.}~\bibnamefont {Kono}}, \bibinfo {author} {\bibfnamefont {A.}~\bibnamefont {Noguchi}}, \bibinfo {author} {\bibfnamefont {K.}~\bibnamefont {Usami}}, \ and\ \bibinfo {author} {\bibfnamefont {Y.}~\bibnamefont {Nakamura}},\ }\bibfield  {title} {\enquote {\bibinfo {title} {{Dissipation-Based Quantum Sensing of Magnons with a Superconducting Qubit}},}\ }\href {\doibase 10.1103/PhysRevLett.125.117701} {\bibfield  {journal} {\bibinfo  {journal} {Phys. Rev. Lett.}\ }\textbf {\bibinfo {volume} {125}},\ \bibinfo {pages} {117701} (\bibinfo {year} {2020})}\BibitemShut {NoStop}%
\bibitem [{\citenamefont {Xu}\ \emph {et~al.}(2023)\citenamefont {Xu}, \citenamefont {Gu}, \citenamefont {Li}, \citenamefont {Weng}, \citenamefont {Wang}, \citenamefont {Li}, \citenamefont {Wang}, \citenamefont {Zhu},\ and\ \citenamefont {You}}]{Xu2023May}%
  \BibitemOpen
  \bibfield  {author} {\bibinfo {author} {\bibfnamefont {D.}~\bibnamefont {Xu}}, \bibinfo {author} {\bibfnamefont {X.-K.}\ \bibnamefont {Gu}}, \bibinfo {author} {\bibfnamefont {H.-K.}\ \bibnamefont {Li}}, \bibinfo {author} {\bibfnamefont {Y.-C.}\ \bibnamefont {Weng}}, \bibinfo {author} {\bibfnamefont {Y.-P.}\ \bibnamefont {Wang}}, \bibinfo {author} {\bibfnamefont {J.}~\bibnamefont {Li}}, \bibinfo {author} {\bibfnamefont {H.}~\bibnamefont {Wang}}, \bibinfo {author} {\bibfnamefont {S.-Y.}\ \bibnamefont {Zhu}}, \ and\ \bibinfo {author} {\bibfnamefont {J.~Q.}\ \bibnamefont {You}},\ }\bibfield  {title} {\enquote {\bibinfo {title} {{Quantum Control of a Single Magnon in a Macroscopic Spin System}},}\ }\href {\doibase 10.1103/PhysRevLett.130.193603} {\bibfield  {journal} {\bibinfo  {journal} {Phys. Rev. Lett.}\ }\textbf {\bibinfo {volume} {130}},\ \bibinfo {pages} {193603} (\bibinfo {year} {2023})}\BibitemShut {NoStop}%
\bibitem [{\citenamefont {Crescini}\ \emph {et~al.}(2020)\citenamefont {Crescini}, \citenamefont {Alesini}, \citenamefont {Braggio}, \citenamefont {Carugno}, \citenamefont {D'Agostino}, \citenamefont {Di~Gioacchino}, \citenamefont {Falferi}, \citenamefont {Gambardella}, \citenamefont {Gatti}, \citenamefont {Iannone}, \citenamefont {Ligi}, \citenamefont {Lombardi}, \citenamefont {Ortolan}, \citenamefont {Pengo}, \citenamefont {Ruoso},\ and\ \citenamefont {Taffarello}}]{Crescini_PRL_2020}%
  \BibitemOpen
  \bibfield  {author} {\bibinfo {author} {\bibfnamefont {N.}~\bibnamefont {Crescini}}, \bibinfo {author} {\bibfnamefont {D.}~\bibnamefont {Alesini}}, \bibinfo {author} {\bibfnamefont {C.}~\bibnamefont {Braggio}}, \bibinfo {author} {\bibfnamefont {G.}~\bibnamefont {Carugno}}, \bibinfo {author} {\bibfnamefont {D.}~\bibnamefont {D'Agostino}}, \bibinfo {author} {\bibfnamefont {D.}~\bibnamefont {Di~Gioacchino}}, \bibinfo {author} {\bibfnamefont {P.}~\bibnamefont {Falferi}}, \bibinfo {author} {\bibfnamefont {U.}~\bibnamefont {Gambardella}}, \bibinfo {author} {\bibfnamefont {C.}~\bibnamefont {Gatti}}, \bibinfo {author} {\bibfnamefont {G.}~\bibnamefont {Iannone}}, \bibinfo {author} {\bibfnamefont {C.}~\bibnamefont {Ligi}}, \bibinfo {author} {\bibfnamefont {A.}~\bibnamefont {Lombardi}}, \bibinfo {author} {\bibfnamefont {A.}~\bibnamefont {Ortolan}}, \bibinfo {author} {\bibfnamefont {R.}~\bibnamefont {Pengo}}, \bibinfo {author} {\bibfnamefont {G.}~\bibnamefont {Ruoso}}, \ and\ \bibinfo {author} {\bibfnamefont
  {L.}~\bibnamefont {Taffarello}} (\bibinfo {collaboration} {QUAX Collaboration}),\ }\bibfield  {title} {\enquote {\bibinfo {title} {Axion search with a quantum-limited ferromagnetic haloscope},}\ }\href {\doibase 10.1103/PhysRevLett.124.171801} {\bibfield  {journal} {\bibinfo  {journal} {Phys. Rev. Lett.}\ }\textbf {\bibinfo {volume} {124}},\ \bibinfo {pages} {171801} (\bibinfo {year} {2020})}\BibitemShut {NoStop}%
\bibitem [{\citenamefont {Flower}\ \emph {et~al.}(2019)\citenamefont {Flower}, \citenamefont {Bourhill}, \citenamefont {Goryachev},\ and\ \citenamefont {Tobar}}]{Flower_PhysDarkUniv_2019}%
  \BibitemOpen
  \bibfield  {author} {\bibinfo {author} {\bibfnamefont {G.}~\bibnamefont {Flower}}, \bibinfo {author} {\bibfnamefont {J.}~\bibnamefont {Bourhill}}, \bibinfo {author} {\bibfnamefont {M.}~\bibnamefont {Goryachev}}, \ and\ \bibinfo {author} {\bibfnamefont {M.~E.}\ \bibnamefont {Tobar}},\ }\bibfield  {title} {\enquote {\bibinfo {title} {Broadening frequency range of a ferromagnetic axion haloscope with strongly coupled cavity--magnon polaritons},}\ }\href@noop {} {\bibfield  {journal} {\bibinfo  {journal} {Physics of the Dark Universe}\ }\textbf {\bibinfo {volume} {25}},\ \bibinfo {pages} {100306} (\bibinfo {year} {2019})}\BibitemShut {NoStop}%
\bibitem [{\citenamefont {Petrenko}\ \emph {et~al.}(1997)\citenamefont {Petrenko}, \citenamefont {Ritter}, \citenamefont {Yethiraj},\ and\ \citenamefont {Paul}}]{Petrenko1997Dec}%
  \BibitemOpen
  \bibfield  {author} {\bibinfo {author} {\bibfnamefont {O.~A.}\ \bibnamefont {Petrenko}}, \bibinfo {author} {\bibfnamefont {C.}~\bibnamefont {Ritter}}, \bibinfo {author} {\bibfnamefont {M.}~\bibnamefont {Yethiraj}}, \ and\ \bibinfo {author} {\bibfnamefont {D.~M.}\ \bibnamefont {Paul}},\ }\bibfield  {title} {\enquote {\bibinfo {title} {{Spin-liquid behavior of the gadolinium gallium garnet}},}\ }\href {\doibase 10.1016/S0921-4526(97)00705-9} {\bibfield  {journal} {\bibinfo  {journal} {Physica B}\ }\textbf {\bibinfo {volume} {241-243}},\ \bibinfo {pages} {727--729} (\bibinfo {year} {1997})}\BibitemShut {NoStop}%
\bibitem [{\citenamefont {Knauer}\ \emph {et~al.}(2023)\citenamefont {Knauer}, \citenamefont {Dav{\ifmmode\acute{\imath}\else\'{\i}\fi}dkov{\ifmmode\acute{a}\else\'{a}\fi}}, \citenamefont {Schmoll}, \citenamefont {Serha}, \citenamefont {Voronov}, \citenamefont {Wang}, \citenamefont {Verba}, \citenamefont {Dobrovolskiy}, \citenamefont {Lindner}, \citenamefont {Reimann}, \citenamefont {Dubs}, \citenamefont {Urb{\ifmmode\acute{a}\else\'{a}\fi}nek},\ and\ \citenamefont {Chumak}}]{Knauer2023Apr_JAP}%
  \BibitemOpen
  \bibfield  {author} {\bibinfo {author} {\bibfnamefont {S.}~\bibnamefont {Knauer}}, \bibinfo {author} {\bibfnamefont {K.}~\bibnamefont {Dav{\ifmmode\acute{\imath}\else\'{\i}\fi}dkov{\ifmmode\acute{a}\else\'{a}\fi}}}, \bibinfo {author} {\bibfnamefont {D.}~\bibnamefont {Schmoll}}, \bibinfo {author} {\bibfnamefont {R.~O.}\ \bibnamefont {Serha}}, \bibinfo {author} {\bibfnamefont {A.}~\bibnamefont {Voronov}}, \bibinfo {author} {\bibfnamefont {Q.}~\bibnamefont {Wang}}, \bibinfo {author} {\bibfnamefont {R.}~\bibnamefont {Verba}}, \bibinfo {author} {\bibfnamefont {O.~V.}\ \bibnamefont {Dobrovolskiy}}, \bibinfo {author} {\bibfnamefont {M.}~\bibnamefont {Lindner}}, \bibinfo {author} {\bibfnamefont {T.}~\bibnamefont {Reimann}}, \bibinfo {author} {\bibfnamefont {C.}~\bibnamefont {Dubs}}, \bibinfo {author} {\bibfnamefont {M.}~\bibnamefont {Urb{\ifmmode\acute{a}\else\'{a}\fi}nek}}, \ and\ \bibinfo {author} {\bibfnamefont {A.~V.}\ \bibnamefont {Chumak}},\ }\bibfield  {title} {\enquote {\bibinfo {title} {{Propagating
  spin-wave spectroscopy in a liquid-phase epitaxial nanometer-thick YIG film at millikelvin temperatures}},}\ }\href {\doibase 10.1063/5.0137437} {\bibfield  {journal} {\bibinfo  {journal} {J. Appl. Phys.}\ }\textbf {\bibinfo {volume} {133}} (\bibinfo {year} {2023}),\ 10.1063/5.0137437}\BibitemShut {NoStop}%
\bibitem [{\citenamefont {Li}\ \emph {et~al.}(2019)\citenamefont {Li}, \citenamefont {Polakovic}, \citenamefont {Wang}, \citenamefont {Xu}, \citenamefont {Lendinez}, \citenamefont {Zhang}, \citenamefont {Ding}, \citenamefont {Khaire}, \citenamefont {Saglam}, \citenamefont {Divan}, \citenamefont {Pearson}, \citenamefont {Kwok}, \citenamefont {Xiao}, \citenamefont {Novosad}, \citenamefont {Hoffmann},\ and\ \citenamefont {Zhang}}]{Li2019Sep}%
  \BibitemOpen
  \bibfield  {author} {\bibinfo {author} {\bibfnamefont {Y.}~\bibnamefont {Li}}, \bibinfo {author} {\bibfnamefont {T.}~\bibnamefont {Polakovic}}, \bibinfo {author} {\bibfnamefont {Y.-L.}\ \bibnamefont {Wang}}, \bibinfo {author} {\bibfnamefont {J.}~\bibnamefont {Xu}}, \bibinfo {author} {\bibfnamefont {S.}~\bibnamefont {Lendinez}}, \bibinfo {author} {\bibfnamefont {Z.}~\bibnamefont {Zhang}}, \bibinfo {author} {\bibfnamefont {J.}~\bibnamefont {Ding}}, \bibinfo {author} {\bibfnamefont {T.}~\bibnamefont {Khaire}}, \bibinfo {author} {\bibfnamefont {H.}~\bibnamefont {Saglam}}, \bibinfo {author} {\bibfnamefont {R.}~\bibnamefont {Divan}}, \bibinfo {author} {\bibfnamefont {J.}~\bibnamefont {Pearson}}, \bibinfo {author} {\bibfnamefont {W.-K.}\ \bibnamefont {Kwok}}, \bibinfo {author} {\bibfnamefont {Z.}~\bibnamefont {Xiao}}, \bibinfo {author} {\bibfnamefont {V.}~\bibnamefont {Novosad}}, \bibinfo {author} {\bibfnamefont {A.}~\bibnamefont {Hoffmann}}, \ and\ \bibinfo {author} {\bibfnamefont {W.}~\bibnamefont {Zhang}},\
  }\bibfield  {title} {\enquote {\bibinfo {title} {{Strong Coupling between Magnons and Microwave Photons in On-Chip Ferromagnet-Superconductor Thin-Film Devices}},}\ }\href {\doibase 10.1103/PhysRevLett.123.107701} {\bibfield  {journal} {\bibinfo  {journal} {Phys. Rev. Lett.}\ }\textbf {\bibinfo {volume} {123}},\ \bibinfo {pages} {107701} (\bibinfo {year} {2019})}\BibitemShut {NoStop}%
\bibitem [{\citenamefont {Hou}\ and\ \citenamefont {Liu}(2019)}]{Hou2019Sep}%
  \BibitemOpen
  \bibfield  {author} {\bibinfo {author} {\bibfnamefont {J.~T.}\ \bibnamefont {Hou}}\ and\ \bibinfo {author} {\bibfnamefont {L.}~\bibnamefont {Liu}},\ }\bibfield  {title} {\enquote {\bibinfo {title} {{Strong Coupling between Microwave Photons and Nanomagnet Magnons}},}\ }\href {\doibase 10.1103/PhysRevLett.123.107702} {\bibfield  {journal} {\bibinfo  {journal} {Phys. Rev. Lett.}\ }\textbf {\bibinfo {volume} {123}},\ \bibinfo {pages} {107702} (\bibinfo {year} {2019})}\BibitemShut {NoStop}%
\bibitem [{\citenamefont {Xu}\ \emph {et~al.}(2021{\natexlab{c}})\citenamefont {Xu}, \citenamefont {Zhong}, \citenamefont {Han}, \citenamefont {Jin}, \citenamefont {Jiang},\ and\ \citenamefont {Zhang}}]{Xu2021Mayprl}%
  \BibitemOpen
  \bibfield  {author} {\bibinfo {author} {\bibfnamefont {J.}~\bibnamefont {Xu}}, \bibinfo {author} {\bibfnamefont {C.}~\bibnamefont {Zhong}}, \bibinfo {author} {\bibfnamefont {X.}~\bibnamefont {Han}}, \bibinfo {author} {\bibfnamefont {D.}~\bibnamefont {Jin}}, \bibinfo {author} {\bibfnamefont {L.}~\bibnamefont {Jiang}}, \ and\ \bibinfo {author} {\bibfnamefont {X.}~\bibnamefont {Zhang}},\ }\bibfield  {title} {\enquote {\bibinfo {title} {{Coherent Gate Operations in Hybrid Magnonics}},}\ }\href {\doibase 10.1103/PhysRevLett.126.207202} {\bibfield  {journal} {\bibinfo  {journal} {Phys. Rev. Lett.}\ }\textbf {\bibinfo {volume} {126}},\ \bibinfo {pages} {207202} (\bibinfo {year} {2021}{\natexlab{c}})}\BibitemShut {NoStop}%
\bibitem [{\citenamefont {Bi}\ \emph {et~al.}(2013)\citenamefont {Bi}, \citenamefont {Hu}, \citenamefont {Jiang}, \citenamefont {Kim}, \citenamefont {Kim}, \citenamefont {Onbasli}, \citenamefont {Dionne},\ and\ \citenamefont {Ross}}]{Bi2013Nov}%
  \BibitemOpen
  \bibfield  {author} {\bibinfo {author} {\bibfnamefont {L.}~\bibnamefont {Bi}}, \bibinfo {author} {\bibfnamefont {J.}~\bibnamefont {Hu}}, \bibinfo {author} {\bibfnamefont {P.}~\bibnamefont {Jiang}}, \bibinfo {author} {\bibfnamefont {H.~S.}\ \bibnamefont {Kim}}, \bibinfo {author} {\bibfnamefont {D.~H.}\ \bibnamefont {Kim}}, \bibinfo {author} {\bibfnamefont {M.~C.}\ \bibnamefont {Onbasli}}, \bibinfo {author} {\bibfnamefont {G.~F.}\ \bibnamefont {Dionne}}, \ and\ \bibinfo {author} {\bibfnamefont {C.~A.}\ \bibnamefont {Ross}},\ }\bibfield  {title} {\enquote {\bibinfo {title} {{Magneto-Optical Thin Films for On-Chip Monolithic Integration of Non-Reciprocal Photonic Devices}},}\ }\href {\doibase 10.3390/ma6115094} {\bibfield  {journal} {\bibinfo  {journal} {Materials}\ }\textbf {\bibinfo {volume} {6}},\ \bibinfo {pages} {5094--5117} (\bibinfo {year} {2013})}\BibitemShut {NoStop}%
\bibitem [{\citenamefont {Onbasli}\ \emph {et~al.}(2014)\citenamefont {Onbasli}, \citenamefont {Goto}, \citenamefont {Sun}, \citenamefont {Huynh},\ and\ \citenamefont {Ross}}]{Onbasli2014Oct}%
  \BibitemOpen
  \bibfield  {author} {\bibinfo {author} {\bibfnamefont {M.~C.}\ \bibnamefont {Onbasli}}, \bibinfo {author} {\bibfnamefont {T.}~\bibnamefont {Goto}}, \bibinfo {author} {\bibfnamefont {X.}~\bibnamefont {Sun}}, \bibinfo {author} {\bibfnamefont {N.}~\bibnamefont {Huynh}}, \ and\ \bibinfo {author} {\bibfnamefont {C.~A.}\ \bibnamefont {Ross}},\ }\bibfield  {title} {\enquote {\bibinfo {title} {{Integration of bulk-quality thin film magneto-optical cerium-doped yttrium iron garnet on silicon nitride photonic substrates}},}\ }\href {\doibase 10.1364/OE.22.025183} {\bibfield  {journal} {\bibinfo  {journal} {Opt. Express}\ }\textbf {\bibinfo {volume} {22}},\ \bibinfo {pages} {25183--25192} (\bibinfo {year} {2014})}\BibitemShut {NoStop}%
\bibitem [{\citenamefont {Guo}\ \emph {et~al.}(2022)\citenamefont {Guo}, \citenamefont {McCullian}, \citenamefont {Chris~Hammel},\ and\ \citenamefont {Yang}}]{Guo2022Nov}%
  \BibitemOpen
  \bibfield  {author} {\bibinfo {author} {\bibfnamefont {S.}~\bibnamefont {Guo}}, \bibinfo {author} {\bibfnamefont {B.}~\bibnamefont {McCullian}}, \bibinfo {author} {\bibfnamefont {P.}~\bibnamefont {Chris~Hammel}}, \ and\ \bibinfo {author} {\bibfnamefont {F.}~\bibnamefont {Yang}},\ }\bibfield  {title} {\enquote {\bibinfo {title} {{Low damping at few-K temperatures in Y3Fe5O12 epitaxial films isolated from Gd3Ga5O12 substrate using a diamagnetic Y3Sc2.5Al2.5O12 spacer}},}\ }\href {\doibase 10.1016/j.jmmm.2022.169795} {\bibfield  {journal} {\bibinfo  {journal} {J. Magn. Magn. Mater.}\ }\textbf {\bibinfo {volume} {562}},\ \bibinfo {pages} {169795} (\bibinfo {year} {2022})}\BibitemShut {NoStop}%
\bibitem [{\citenamefont {Guo}\ \emph {et~al.}(2023)\citenamefont {Guo}, \citenamefont {Russell}, \citenamefont {Lanier}, \citenamefont {Da}, \citenamefont {Hammel},\ and\ \citenamefont {Yang}}]{Guo2023Jun}%
  \BibitemOpen
  \bibfield  {author} {\bibinfo {author} {\bibfnamefont {S.}~\bibnamefont {Guo}}, \bibinfo {author} {\bibfnamefont {D.}~\bibnamefont {Russell}}, \bibinfo {author} {\bibfnamefont {J.}~\bibnamefont {Lanier}}, \bibinfo {author} {\bibfnamefont {H.}~\bibnamefont {Da}}, \bibinfo {author} {\bibfnamefont {P.~C.}\ \bibnamefont {Hammel}}, \ and\ \bibinfo {author} {\bibfnamefont {F.}~\bibnamefont {Yang}},\ }\bibfield  {title} {\enquote {\bibinfo {title} {{Strong on-Chip Microwave Photon{\textendash}Magnon Coupling Using Ultralow-Damping Epitaxial Y3Fe5O12 Films at 2 K}},}\ }\href {\doibase 10.1021/acs.nanolett.3c00959} {\bibfield  {journal} {\bibinfo  {journal} {Nano Lett.}\ }\textbf {\bibinfo {volume} {23}},\ \bibinfo {pages} {5055--5060} (\bibinfo {year} {2023})}\BibitemShut {NoStop}%
\bibitem [{\citenamefont {Levy}\ \emph {et~al.}(1998{\natexlab{a}})\citenamefont {Levy}, \citenamefont {Osgood}, \citenamefont {Kumar},\ and\ \citenamefont {Bakhru}}]{Levy1998Jun}%
  \BibitemOpen
  \bibfield  {author} {\bibinfo {author} {\bibfnamefont {M.}~\bibnamefont {Levy}}, \bibinfo {author} {\bibfnamefont {R.~M.}\ \bibnamefont {Osgood}}, \bibinfo {author} {\bibfnamefont {A.}~\bibnamefont {Kumar}}, \ and\ \bibinfo {author} {\bibfnamefont {H.}~\bibnamefont {Bakhru}},\ }\bibfield  {title} {\enquote {\bibinfo {title} {{Crystal ion slicing of single-crystal magnetic garnet films}},}\ }\href {\doibase 10.1063/1.367673} {\bibfield  {journal} {\bibinfo  {journal} {J. Appl. Phys.}\ }\textbf {\bibinfo {volume} {83}},\ \bibinfo {pages} {6759--6761} (\bibinfo {year} {1998}{\natexlab{a}})}\BibitemShut {NoStop}%
\bibitem [{\citenamefont {Heyroth}\ \emph {et~al.}(2019)\citenamefont {Heyroth}, \citenamefont {Hauser}, \citenamefont {Trempler}, \citenamefont {Geyer}, \citenamefont {Syrowatka}, \citenamefont {Dreyer}, \citenamefont {Ebbinghaus}, \citenamefont {Woltersdorf},\ and\ \citenamefont {Schmidt}}]{Heyroth2019Nov}%
  \BibitemOpen
  \bibfield  {author} {\bibinfo {author} {\bibfnamefont {F.}~\bibnamefont {Heyroth}}, \bibinfo {author} {\bibfnamefont {C.}~\bibnamefont {Hauser}}, \bibinfo {author} {\bibfnamefont {P.}~\bibnamefont {Trempler}}, \bibinfo {author} {\bibfnamefont {P.}~\bibnamefont {Geyer}}, \bibinfo {author} {\bibfnamefont {F.}~\bibnamefont {Syrowatka}}, \bibinfo {author} {\bibfnamefont {R.}~\bibnamefont {Dreyer}}, \bibinfo {author} {\bibfnamefont {S.~G.}\ \bibnamefont {Ebbinghaus}}, \bibinfo {author} {\bibfnamefont {G.}~\bibnamefont {Woltersdorf}}, \ and\ \bibinfo {author} {\bibfnamefont {G.}~\bibnamefont {Schmidt}},\ }\bibfield  {title} {\enquote {\bibinfo {title} {{Monocrystalline Freestanding Three-Dimensional Yttrium-Iron-Garnet Magnon Nanoresonators}},}\ }\href {\doibase 10.1103/PhysRevApplied.12.054031} {\bibfield  {journal} {\bibinfo  {journal} {Phys. Rev. Appl.}\ }\textbf {\bibinfo {volume} {12}},\ \bibinfo {pages} {054031} (\bibinfo {year} {2019})}\BibitemShut {NoStop}%
\bibitem [{\citenamefont {Haigh}, \citenamefont {Chakalov},\ and\ \citenamefont {Ramsay}(2020)}]{Haigh2020Oct}%
  \BibitemOpen
  \bibfield  {author} {\bibinfo {author} {\bibfnamefont {J.~A.}\ \bibnamefont {Haigh}}, \bibinfo {author} {\bibfnamefont {R.~A.}\ \bibnamefont {Chakalov}}, \ and\ \bibinfo {author} {\bibfnamefont {A.~J.}\ \bibnamefont {Ramsay}},\ }\bibfield  {title} {\enquote {\bibinfo {title} {{Subpicoliter Magnetoptical Cavities}},}\ }\href {\doibase 10.1103/PhysRevApplied.14.044005} {\bibfield  {journal} {\bibinfo  {journal} {Phys. Rev. Appl.}\ }\textbf {\bibinfo {volume} {14}},\ \bibinfo {pages} {044005} (\bibinfo {year} {2020})}\BibitemShut {NoStop}%
\bibitem [{\citenamefont {Baity}\ \emph {et~al.}(2021)\citenamefont {Baity}, \citenamefont {Bozhko}, \citenamefont {Mac{\ifmmode\hat{e}\else\^{e}\fi}do}, \citenamefont {Smith}, \citenamefont {Holland}, \citenamefont {Danilin}, \citenamefont {Seferai}, \citenamefont {Barbosa}, \citenamefont {Peroor}, \citenamefont {Goldman}, \citenamefont {Nasti}, \citenamefont {Paul}, \citenamefont {Hadfield}, \citenamefont {McVitie},\ and\ \citenamefont {Weides}}]{Baity2021Jul}%
  \BibitemOpen
  \bibfield  {author} {\bibinfo {author} {\bibfnamefont {P.~G.}\ \bibnamefont {Baity}}, \bibinfo {author} {\bibfnamefont {D.~A.}\ \bibnamefont {Bozhko}}, \bibinfo {author} {\bibfnamefont {R.}~\bibnamefont {Mac{\ifmmode\hat{e}\else\^{e}\fi}do}}, \bibinfo {author} {\bibfnamefont {W.}~\bibnamefont {Smith}}, \bibinfo {author} {\bibfnamefont {R.~C.}\ \bibnamefont {Holland}}, \bibinfo {author} {\bibfnamefont {S.}~\bibnamefont {Danilin}}, \bibinfo {author} {\bibfnamefont {V.}~\bibnamefont {Seferai}}, \bibinfo {author} {\bibfnamefont {J.}~\bibnamefont {Barbosa}}, \bibinfo {author} {\bibfnamefont {R.~R.}\ \bibnamefont {Peroor}}, \bibinfo {author} {\bibfnamefont {S.}~\bibnamefont {Goldman}}, \bibinfo {author} {\bibfnamefont {U.}~\bibnamefont {Nasti}}, \bibinfo {author} {\bibfnamefont {J.}~\bibnamefont {Paul}}, \bibinfo {author} {\bibfnamefont {R.~H.}\ \bibnamefont {Hadfield}}, \bibinfo {author} {\bibfnamefont {S.}~\bibnamefont {McVitie}}, \ and\ \bibinfo {author} {\bibfnamefont {M.}~\bibnamefont {Weides}},\ }\bibfield
  {title} {\enquote {\bibinfo {title} {{Strong magnon{\textendash}photon coupling with chip-integrated YIG in the zero-temperature limit}},}\ }\href {\doibase 10.1063/5.0054837} {\bibfield  {journal} {\bibinfo  {journal} {Appl. Phys. Lett.}\ }\textbf {\bibinfo {volume} {119}} (\bibinfo {year} {2021}),\ 10.1063/5.0054837}\BibitemShut {NoStop}%
\bibitem [{\citenamefont {Bedell}\ \emph {et~al.}(2012)\citenamefont {Bedell}, \citenamefont {Shahrjerdi}, \citenamefont {Hekmatshoar}, \citenamefont {Fogel}, \citenamefont {Lauro}, \citenamefont {Ott}, \citenamefont {Sosa},\ and\ \citenamefont {Sadana}}]{Bedell2012Feb}%
  \BibitemOpen
  \bibfield  {author} {\bibinfo {author} {\bibfnamefont {S.~W.}\ \bibnamefont {Bedell}}, \bibinfo {author} {\bibfnamefont {D.}~\bibnamefont {Shahrjerdi}}, \bibinfo {author} {\bibfnamefont {B.}~\bibnamefont {Hekmatshoar}}, \bibinfo {author} {\bibfnamefont {K.}~\bibnamefont {Fogel}}, \bibinfo {author} {\bibfnamefont {P.~A.}\ \bibnamefont {Lauro}}, \bibinfo {author} {\bibfnamefont {J.~A.}\ \bibnamefont {Ott}}, \bibinfo {author} {\bibfnamefont {N.}~\bibnamefont {Sosa}}, \ and\ \bibinfo {author} {\bibfnamefont {D.}~\bibnamefont {Sadana}},\ }\bibfield  {title} {\enquote {\bibinfo {title} {{Kerf-Less Removal of Si, Ge, and III{\textendash}V Layers by Controlled Spalling to Enable Low-Cost PV Technologies}},}\ }\href {\doibase 10.1109/JPHOTOV.2012.2184267} {\bibfield  {journal} {\bibinfo  {journal} {IEEE J. Photovoltaics}\ }\textbf {\bibinfo {volume} {2}},\ \bibinfo {pages} {141--147} (\bibinfo {year} {2012})}\BibitemShut {NoStop}%
\bibitem [{\citenamefont {Bedell}\ \emph {et~al.}(2017)\citenamefont {Bedell}, \citenamefont {Lauro}, \citenamefont {Ott}, \citenamefont {Fogel},\ and\ \citenamefont {Sadana}}]{Bedell2017Jul}%
  \BibitemOpen
  \bibfield  {author} {\bibinfo {author} {\bibfnamefont {S.~W.}\ \bibnamefont {Bedell}}, \bibinfo {author} {\bibfnamefont {P.}~\bibnamefont {Lauro}}, \bibinfo {author} {\bibfnamefont {J.~A.}\ \bibnamefont {Ott}}, \bibinfo {author} {\bibfnamefont {K.}~\bibnamefont {Fogel}}, \ and\ \bibinfo {author} {\bibfnamefont {D.~K.}\ \bibnamefont {Sadana}},\ }\bibfield  {title} {\enquote {\bibinfo {title} {{Layer transfer of bulk gallium nitride by controlled spalling}},}\ }\href {\doibase 10.1063/1.4986646} {\bibfield  {journal} {\bibinfo  {journal} {J. Appl. Phys.}\ }\textbf {\bibinfo {volume} {122}} (\bibinfo {year} {2017}),\ 10.1063/1.4986646}\BibitemShut {NoStop}%
\bibitem [{\citenamefont {Horn}\ \emph {et~al.}(2024)\citenamefont {Horn}, \citenamefont {Wicker}, \citenamefont {Wellisz}, \citenamefont {Zeledon}, \citenamefont {Nittala}, \citenamefont {Heremans}, \citenamefont {Awschalom},\ and\ \citenamefont {Guha}}]{Horn2024Oct}%
  \BibitemOpen
  \bibfield  {author} {\bibinfo {author} {\bibfnamefont {C.~P.}\ \bibnamefont {Horn}}, \bibinfo {author} {\bibfnamefont {C.}~\bibnamefont {Wicker}}, \bibinfo {author} {\bibfnamefont {A.}~\bibnamefont {Wellisz}}, \bibinfo {author} {\bibfnamefont {C.}~\bibnamefont {Zeledon}}, \bibinfo {author} {\bibfnamefont {P.~V.~K.}\ \bibnamefont {Nittala}}, \bibinfo {author} {\bibfnamefont {F.~J.}\ \bibnamefont {Heremans}}, \bibinfo {author} {\bibfnamefont {D.~D.}\ \bibnamefont {Awschalom}}, \ and\ \bibinfo {author} {\bibfnamefont {S.}~\bibnamefont {Guha}},\ }\bibfield  {title} {\enquote {\bibinfo {title} {{Controlled Spalling of 4H Silicon Carbide with Investigated Spin Coherence for Quantum Engineering Integration}},}\ }\href {\doibase 10.1021/acsnano.4c10978} {\bibfield  {journal} {\bibinfo  {journal} {ACS Nano}\ }\textbf {\bibinfo {volume} {2024}} (\bibinfo {year} {2024}),\ 10.1021/acsnano.4c10978}\BibitemShut {NoStop}%
\bibitem [{\citenamefont {Suo}\ and\ \citenamefont {Hutchinson}(1989)}]{Suo1989Jan}%
  \BibitemOpen
  \bibfield  {author} {\bibinfo {author} {\bibfnamefont {Z.}~\bibnamefont {Suo}}\ and\ \bibinfo {author} {\bibfnamefont {J.~W.}\ \bibnamefont {Hutchinson}},\ }\bibfield  {title} {\enquote {\bibinfo {title} {{Steady-state cracking in brittle substrates beneath adherent films}},}\ }\href {\doibase 10.1016/0020-7683(89)90096-6} {\bibfield  {journal} {\bibinfo  {journal} {Int. J. Solids Struct.}\ }\textbf {\bibinfo {volume} {25}},\ \bibinfo {pages} {1337--1353} (\bibinfo {year} {1989})}\BibitemShut {NoStop}%
\bibitem [{\citenamefont {Levy}\ \emph {et~al.}(1998{\natexlab{b}})\citenamefont {Levy}, \citenamefont {Osgood}, \citenamefont {Liu}, \citenamefont {Cross}, \citenamefont {Cargill}, \citenamefont {Kumar},\ and\ \citenamefont {Bakhru}}]{Levy1998Oct}%
  \BibitemOpen
  \bibfield  {author} {\bibinfo {author} {\bibfnamefont {M.}~\bibnamefont {Levy}}, \bibinfo {author} {\bibfnamefont {R.~M.}\ \bibnamefont {Osgood}}, \bibinfo {author} {\bibfnamefont {R.}~\bibnamefont {Liu}}, \bibinfo {author} {\bibfnamefont {L.~E.}\ \bibnamefont {Cross}}, \bibinfo {author} {\bibfnamefont {G.~S.}\ \bibnamefont {Cargill}}, \bibinfo {author} {\bibfnamefont {A.}~\bibnamefont {Kumar}}, \ and\ \bibinfo {author} {\bibfnamefont {H.}~\bibnamefont {Bakhru}},\ }\bibfield  {title} {\enquote {\bibinfo {title} {{Fabrication of single-crystal lithium niobate films by crystal ion slicing}},}\ }\href {\doibase 10.1063/1.121801} {\bibfield  {journal} {\bibinfo  {journal} {Appl. Phys. Lett.}\ }\textbf {\bibinfo {volume} {73}},\ \bibinfo {pages} {2293--2295} (\bibinfo {year} {1998}{\natexlab{b}})}\BibitemShut {NoStop}%
\bibitem [{\citenamefont {Dubs}\ \emph {et~al.}(2017)\citenamefont {Dubs}, \citenamefont {Surzhenko}, \citenamefont {Linke}, \citenamefont {Danilewsky}, \citenamefont {Br{\ifmmode\ddot{u}\else\"{u}\fi}ckner},\ and\ \citenamefont {Dellith}}]{Dubs2017Apr_JPD}%
  \BibitemOpen
  \bibfield  {author} {\bibinfo {author} {\bibfnamefont {C.}~\bibnamefont {Dubs}}, \bibinfo {author} {\bibfnamefont {O.}~\bibnamefont {Surzhenko}}, \bibinfo {author} {\bibfnamefont {R.}~\bibnamefont {Linke}}, \bibinfo {author} {\bibfnamefont {A.}~\bibnamefont {Danilewsky}}, \bibinfo {author} {\bibfnamefont {U.}~\bibnamefont {Br{\ifmmode\ddot{u}\else\"{u}\fi}ckner}}, \ and\ \bibinfo {author} {\bibfnamefont {J.}~\bibnamefont {Dellith}},\ }\bibfield  {title} {\enquote {\bibinfo {title} {{Sub-micrometer yttrium iron garnet LPE films with low ferromagnetic resonance losses}},}\ }\href {\doibase 10.1088/1361-6463/aa6b1c} {\bibfield  {journal} {\bibinfo  {journal} {J. Phys. D: Appl. Phys.}\ }\textbf {\bibinfo {volume} {50}},\ \bibinfo {pages} {204005} (\bibinfo {year} {2017})}\BibitemShut {NoStop}%
\bibitem [{\citenamefont {Dubs}\ \emph {et~al.}(2020)\citenamefont {Dubs}, \citenamefont {Surzhenko}, \citenamefont {Thomas}, \citenamefont {Osten}, \citenamefont {Schneider}, \citenamefont {Lenz}, \citenamefont {Grenzer}, \citenamefont {H{\ifmmode\ddot{u}\else\"{u}\fi}bner},\ and\ \citenamefont {Wendler}}]{Dubs2020Feb_PRM}%
  \BibitemOpen
  \bibfield  {author} {\bibinfo {author} {\bibfnamefont {C.}~\bibnamefont {Dubs}}, \bibinfo {author} {\bibfnamefont {O.}~\bibnamefont {Surzhenko}}, \bibinfo {author} {\bibfnamefont {R.}~\bibnamefont {Thomas}}, \bibinfo {author} {\bibfnamefont {J.}~\bibnamefont {Osten}}, \bibinfo {author} {\bibfnamefont {T.}~\bibnamefont {Schneider}}, \bibinfo {author} {\bibfnamefont {K.}~\bibnamefont {Lenz}}, \bibinfo {author} {\bibfnamefont {J.}~\bibnamefont {Grenzer}}, \bibinfo {author} {\bibfnamefont {R.}~\bibnamefont {H{\ifmmode\ddot{u}\else\"{u}\fi}bner}}, \ and\ \bibinfo {author} {\bibfnamefont {E.}~\bibnamefont {Wendler}},\ }\bibfield  {title} {\enquote {\bibinfo {title} {{Low damping and microstructural perfection of sub-40nm-thin yttrium iron garnet films grown by liquid phase epitaxy}},}\ }\href {\doibase 10.1103/PhysRevMaterials.4.024416} {\bibfield  {journal} {\bibinfo  {journal} {Phys. Rev. Mater.}\ }\textbf {\bibinfo {volume} {4}},\ \bibinfo {pages} {024416} (\bibinfo {year} {2020})}\BibitemShut {NoStop}%
\bibitem [{\citenamefont {Pirro}\ \emph {et~al.}(2014)\citenamefont {Pirro}, \citenamefont {Br{\ifmmode\ddot{a}\else\"{a}\fi}cher}, \citenamefont {Chumak}, \citenamefont {L{\ifmmode\ddot{a}\else\"{a}\fi}gel}, \citenamefont {Dubs}, \citenamefont {Surzhenko}, \citenamefont {G{\ifmmode\ddot{o}\else\"{o}\fi}rnert}, \citenamefont {Leven},\ and\ \citenamefont {Hillebrands}}]{Pirro2014Jan_APL}%
  \BibitemOpen
  \bibfield  {author} {\bibinfo {author} {\bibfnamefont {P.}~\bibnamefont {Pirro}}, \bibinfo {author} {\bibfnamefont {T.}~\bibnamefont {Br{\ifmmode\ddot{a}\else\"{a}\fi}cher}}, \bibinfo {author} {\bibfnamefont {A.~V.}\ \bibnamefont {Chumak}}, \bibinfo {author} {\bibfnamefont {B.}~\bibnamefont {L{\ifmmode\ddot{a}\else\"{a}\fi}gel}}, \bibinfo {author} {\bibfnamefont {C.}~\bibnamefont {Dubs}}, \bibinfo {author} {\bibfnamefont {O.}~\bibnamefont {Surzhenko}}, \bibinfo {author} {\bibfnamefont {P.}~\bibnamefont {G{\ifmmode\ddot{o}\else\"{o}\fi}rnert}}, \bibinfo {author} {\bibfnamefont {B.}~\bibnamefont {Leven}}, \ and\ \bibinfo {author} {\bibfnamefont {B.}~\bibnamefont {Hillebrands}},\ }\bibfield  {title} {\enquote {\bibinfo {title} {{Spin-wave excitation and propagation in microstructured waveguides of yttrium iron garnet/Pt bilayers}},}\ }\href {\doibase 10.1063/1.4861343} {\bibfield  {journal} {\bibinfo  {journal} {Appl. Phys. Lett.}\ }\textbf {\bibinfo {volume} {104}},\ \bibinfo {pages} {012402} (\bibinfo {year}
  {2014})}\BibitemShut {NoStop}%
\bibitem [{\citenamefont {Boyd}(2008)}]{SM_Boyd2008}%
  \BibitemOpen
  \bibfield  {author} {\bibinfo {author} {\bibfnamefont {R.~W.}\ \bibnamefont {Boyd}},\ }\href {https://www.sciencedirect.com/book/9780123694706/nonlinear-optics#book-description} {\emph {\bibinfo {title} {{Nonlinear Optics}}}}\ (\bibinfo  {publisher} {Elsevier, Academic Press},\ \bibinfo {year} {2008})\BibitemShut {NoStop}%
\bibitem [{SM()}]{SM}%
  \BibitemOpen
  \href@noop {} {\bibinfo  {journal} {Supplemental material}\ }\BibitemShut {NoStop}%
\bibitem [{\citenamefont {Schuster}\ \emph {et~al.}(2010)\citenamefont {Schuster}, \citenamefont {Sears}, \citenamefont {Ginossar}, \citenamefont {DiCarlo}, \citenamefont {Frunzio}, \citenamefont {Morton}, \citenamefont {Wu}, \citenamefont {Briggs}, \citenamefont {Buckley}, \citenamefont {Awschalom},\ and\ \citenamefont {Schoelkopf}}]{Schuster2010Sep}%
  \BibitemOpen
\bibfield  {journal} {  }\bibfield  {author} {\bibinfo {author} {\bibfnamefont {D.~I.}\ \bibnamefont {Schuster}}, \bibinfo {author} {\bibfnamefont {A.~P.}\ \bibnamefont {Sears}}, \bibinfo {author} {\bibfnamefont {E.}~\bibnamefont {Ginossar}}, \bibinfo {author} {\bibfnamefont {L.}~\bibnamefont {DiCarlo}}, \bibinfo {author} {\bibfnamefont {L.}~\bibnamefont {Frunzio}}, \bibinfo {author} {\bibfnamefont {J.~J.~L.}\ \bibnamefont {Morton}}, \bibinfo {author} {\bibfnamefont {H.}~\bibnamefont {Wu}}, \bibinfo {author} {\bibfnamefont {G.~A.~D.}\ \bibnamefont {Briggs}}, \bibinfo {author} {\bibfnamefont {B.~B.}\ \bibnamefont {Buckley}}, \bibinfo {author} {\bibfnamefont {D.~D.}\ \bibnamefont {Awschalom}}, \ and\ \bibinfo {author} {\bibfnamefont {R.~J.}\ \bibnamefont {Schoelkopf}},\ }\bibfield  {title} {\enquote {\bibinfo {title} {{High-Cooperativity Coupling of Electron-Spin Ensembles to Superconducting Cavities}},}\ }\href {\doibase 10.1103/PhysRevLett.105.140501} {\bibfield  {journal} {\bibinfo  {journal} {Phys. Rev.
  Lett.}\ }\textbf {\bibinfo {volume} {105}},\ \bibinfo {pages} {140501} (\bibinfo {year} {2010})}\BibitemShut {NoStop}%
\bibitem [{\citenamefont {Walls}\ and\ \citenamefont {Milburn}(2008)}]{QuantumOptics}%
  \BibitemOpen
  \bibfield  {author} {\bibinfo {author} {\bibfnamefont {D.}~\bibnamefont {Walls}}\ and\ \bibinfo {author} {\bibfnamefont {G.~J.}\ \bibnamefont {Milburn}},\ }\href {https://link.springer.com/book/10.1007/978-3-540-28574-8} {\emph {\bibinfo {title} {{Quantum Optics}}}}\ (\bibinfo  {publisher} {Springer},\ \bibinfo {address} {Berlin, Germany},\ \bibinfo {year} {2008})\BibitemShut {NoStop}%
\bibitem [{\citenamefont {Chen}\ \emph {et~al.}(2022)\citenamefont {Chen}, \citenamefont {Pfeiffer}, \citenamefont {Partanen}, \citenamefont {Fesquet}, \citenamefont {Honasoge}, \citenamefont {Kronowetter}, \citenamefont {Nojiri}, \citenamefont {Renger}, \citenamefont {Fedorov}, \citenamefont {Marx}, \citenamefont {Deppe},\ and\ \citenamefont {Gross}}]{Chen2022Dec}%
  \BibitemOpen
  \bibfield  {author} {\bibinfo {author} {\bibfnamefont {Q.-M.}\ \bibnamefont {Chen}}, \bibinfo {author} {\bibfnamefont {M.}~\bibnamefont {Pfeiffer}}, \bibinfo {author} {\bibfnamefont {M.}~\bibnamefont {Partanen}}, \bibinfo {author} {\bibfnamefont {F.}~\bibnamefont {Fesquet}}, \bibinfo {author} {\bibfnamefont {K.~E.}\ \bibnamefont {Honasoge}}, \bibinfo {author} {\bibfnamefont {F.}~\bibnamefont {Kronowetter}}, \bibinfo {author} {\bibfnamefont {Y.}~\bibnamefont {Nojiri}}, \bibinfo {author} {\bibfnamefont {M.}~\bibnamefont {Renger}}, \bibinfo {author} {\bibfnamefont {K.~G.}\ \bibnamefont {Fedorov}}, \bibinfo {author} {\bibfnamefont {A.}~\bibnamefont {Marx}}, \bibinfo {author} {\bibfnamefont {F.}~\bibnamefont {Deppe}}, \ and\ \bibinfo {author} {\bibfnamefont {R.}~\bibnamefont {Gross}},\ }\bibfield  {title} {\enquote {\bibinfo {title} {{Scattering coefficients of superconducting microwave resonators. I. Transfer matrix approach}},}\ }\href {\doibase 10.1103/PhysRevB.106.214505} {\bibfield  {journal} {\bibinfo
  {journal} {Phys. Rev. B}\ }\textbf {\bibinfo {volume} {106}},\ \bibinfo {pages} {214505} (\bibinfo {year} {2022})}\BibitemShut {NoStop}%
\end{thebibliography}

\providecommand{\noopsort}[1]{}\providecommand{\singleletter}[1]{#1}%

\end{document}